%% file: les_hst_sekimoto_arXiv.tex
\documentclass{jfm}
\usepackage{natbib}
\usepackage{amsmath}
\usepackage{upmath}
\usepackage{amssymb}
\usepackage{amsbsy}
\usepackage{graphicx,color}

\include{set_newcommand}

\title[Vertically localised equilibria in LES of homogeneous shear flow]
{Vertically localised equilibrium solutions in large-eddy simulations of homogeneous shear flow}
\author[ A. Sekimoto and J. Jim\'enez ]
{A\ls T\ls S\ls U\ls S\ls H\ls I\ns S\ls E\ls K\ls I\ls M\ls O\ls T\ls O
  \thanks{Email for correspondence: atsushi.sekimoto@monash.edu.au}\ns
\and\ns 
J\ls A\ls V\ls I\ls E\ls R\ns J\ls I\ls M\ls \'E\ls N\ls E\ls Z}
\affiliation
{
School of Aeronautics, Universidad Polit\'ecnica de Madrid, 28040 Madrid,
Spain\\[\affilskip]
}

\pubyear{}
\volume{}
\pagerange{}
\date{\today}
\begin{document}

\maketitle

\begin{abstract}
Unstable equilibrium solutions in a homogeneous shear flow with sinuous (streamwise-shift-reflection and spanwise-shift-rotation) symmetry
are numerically found in large-eddy simulations (LESes) with no kinetic viscosity.
The small-scale properties are determined by the mixing length scale $l_S$ used to define
eddy viscosity, and the large-scale motion is induced by the mean shear at the integral
scale, which is limited by the spanwise box dimension $L_z$. The fraction $ R_S= L_z/l_S$,
which plays the role of a Reynolds number, is used as a numerical continuation parameter. It
is shown that equilibrium solutions appear by a saddle-node bifurcation as $R_S$ 
increases, and that the flow structures resemble those in plane Couette flow 
with the same sinuous symmetry. 
The vortical structures of both lower- and upper-branch solutions become
spontaneously localised in the vertical direction. 
The lower-branch solution is an edge state at low $R_S$, and takes the form of a thin critical layer as $R_S$ increases, as in
the asymptotic theory of generic shear flow at high-Reynolds numbers. On the other hand, the
upper-branch solutions are characterised by a tall velocity streak with multi-scale multiple
vortical structures. At the higher end of $R_S$, an incipient multiscale structure is found.
The LES turbulence occasionally visits vertically localised states whose vortical structure
resembles the present vertically localised LES equilibria.
\end{abstract}

\section{Introduction}

Nonlinear invariant solutions of the incompressible Navier--Stokes (NS) equations, such as
equilibria~\citep{Nagata1990} or periodic orbits~\citep{KawaharaKida2001}, 
are believed to play an important role in transitional and
self-sustaining turbulence. In particular, it has been proposed that
coherent structures in turbulent flows are incomplete representations of such solutions,
corresponding to times in which the flow approaches an invariant solution in phase
space \citep{jimenez87}. The solutions themselves could then be considered `exact' coherent
structures \citep{Waleffe2001}.
Their properties and significance are reviewed in \cite{KawaharaUhlmannVeen2012}.

An additional advantage of invariant solutions in the description of turbulence is that they
can be exactly reproduced numerically, potentially providing a
well-defined dynamical `alphabet' for the flow evolution, 
and chaotic fluid motions are created by homoclinic or heteroclinic entanglement of their stable/unstable manifolds. 
Their statistical agreement with turbulence at low Reynolds numbers has often been noted 
in the literature \citep{KawaharaKida2001,JimenezKawaharaSimensNagataShiba2005,
	KerswellTutty2007,Viswanath2007}.
Unfortunately, the dynamically important solutions embedded in turbulence
have only been found at low Reynolds numbers in wall-bounded flows such as 
plane Poiseuille or Couette flow ~\citep{KawaharaKida2001,VanVeenKawahara2011,KreilosEckhardt2012,ParkGraham2015}. 
From a practical point of view, they are hard to continue to higher Reynolds numbers
partly because of their increasing complexity and instability, and by the limitations of the numerical resources. 
From the theoretical side, their increasing instability as the Reynolds
number increases calls into question whether the flow would approach them often enough for
them to be considered relevant. 
There has also been a challenge to estimate turbulence statistics by using all of recurrent flows~\citep{ChandlerKerswell2013,Cvitanovic2013}, however, such an exhaustive study is still limited at low Reynolds numbers. A a more promising method will be required to find invariant solutions at high Reynolds numbers with a large number of degrees of freedom, such as the multiple-shooting method~\citep{SanchezNet2010,VanVeenGawaharaMatsumura2011}.
As a consequence they have mostly been discussed in the
context of the transition to turbulence from the laminar state
\citep{SchmiegelEckhardt1997,FaisstEckhardt2003,WedinKerswell2004, WangGibsonWaleffe2007,ItanoGeneralis2009,AvilaMellibovskyRolandHof2013,ZammertEckhardt2015}.

The dynamical system approach on the transition to turbulence advocates that there is an edge state on the basin boundary between the linearly-stable laminar state and turbulent state,
which can be captured by edge tracking~\citep{ItanoToh2001,TohItano2003,SkufcaYorkeEckhardt2006}.
Typically, it determines how the fluid behaves as it transitions from laminar to turbulent states and vice versa. Some of lower-branch solutions often sit on the edge state, and form a critical layer at high Reynolds number~\citep{WangGibsonWaleffe2007,Viswanath2009}.
Such critical layer-type solutions are described by an asymptotic theory called as vortex-wave interaction (VWI)~\citep{HallSmith1991,HallSherwin2010} and their instability has the edge-mode~\citep{DeguchiHall2016}. 
It is shown that vertically-localised equilibrium states can be embedded in any shear flow at high Reynolds number~\citep{BlackburnHallSherwin2013,DeguchiHall2014_PSTA,Deguchi2015}.

The multiscale nature of fully-developed turbulence at high Reynolds number, and the
possible role of coherent structures in the energy cascade raise the question of whether
invariant solutions may also be relevant in such processes. \Citet{VanVeenKidaKawahara2006}
found a periodic orbit in highly symmetric turbulence that results in a $k^{-5/3}$ energy
spectrum, and could be part of the generic energy cascade~\citep{Goto2008}. 
Similar attempts, however, have not been successful in wall-bounded turbulence at high Reynolds number, 
because it is difficult to accommodate the anisotropy and inhomogeneity of the length scales introduced by the wall.

Here, we simplify the problem in two ways. In the first place we substitute shear-driven
wall-bounded flow by a homogeneous shear without walls, 
of which large-scale motion is limited by the spanwise box dimension in a long-term simulation~\citep{SekimotoDongJimenez2016}. 
Secondly, we substitute the full NS equations by large-eddy simulations (LES), of which small-scale property is determined by 
the eddy-viscosity. 

Homogeneous shear turbulence is an idealised case which shares with wall-bounded flows the
basic source of turbulent energy by the shear. Ideal unbounded homogeneous shear flow has no
intrinsic length scale and is believed to grow without bound from any initial condition
\citep{champ:h:c:70,RogersMoin1987,TavoularisKarnik1989,CambonScott1999} but, when simulated
in a finite computational box, it reaches a statistically stationary and homogeneous state
(SS-HST) characterised by repeated bursting reminiscent of near-wall and logarithmic-layer
turbulence \citep{Pumir1996,Gualtieri2007}. The problem was recently revisited by
\cite{SekimotoDongJimenez2016} using direct numerical simulations (DNSes), 
from where the simulation code and parameters in this paper are derived. 
They determined which computational boxes best mimic wall-bounded turbulence,
and showed that the relevant limiting dimension is the spanwise box width. In terms of this
dimension, SS-HST is always minimal in the sense that a single velocity streak tends to
fills the whole span, as in minimal channels \citep{JimenezMoin1991,FloresJimenez2010}. They
showed that this minimal flow, even with no walls, is a very promising model for shear
turbulence with an non-inflectional mean profile, and particularly for the logarithmic layer
of wall-bounded flows.

Large-eddy simulation is based on the idea that the large flow scales, which are explicitly
computed, are essentially independent of the smaller ones, which are typically modelled by
some kind of eddy viscosity \citep{pope00book}. In shear flows, the larger scales usually
account for most of the kinetic energy and momentum transfer, while the smaller ones tend to
be isotropic and basically provide dissipation and act as random perturbations. In
particular, it is known that the large-scale motion is maintained in channels even when the
small scales have been filtered in LES \citep{scov2001,HwangCossu2010}. The key idea of
extending the dynamical system approach to higher Reynolds numbers is to focus on the
quasi-autonomous behaviour of the larger scales by modelling the effect of the smaller ones
using a sub-grid (SG) model 
\citep{YasudaGotoKawahara2014,RawatCossuHwangRincon2015,HwangWillisCossu2016,SasakiKawaharaSekimotoJimenez2016},
where some LES steady states and periodic orbits are numerically obtained in wall-bounded flow 
and they are considered as a representation of large-scale motions.
However, as mentioned above, the anisotropy and inhomogeneity of the length scales
introduced by the wall makes the role of the eddy-viscosity term, which should be zero on the wall, 
on the invariant solutions ambiguous. 
We apply LES here to the identification of equilibrium structures in SS-HST, where the large length scale is constrained and represented by
the spanwise box dimension and the smallest scale is imposed by a SG model, so that the scale separation is statistically homogeneous, 
and the length-scale ratio is used as a grid-independent LES parameter.
The application of LES greatly also decreases the number of degrees of freedom required to track invariant solutions at high Reynolds number.

Section \ref{sec:numer} presents the numerical method and establishes a
baseline LES with sinuous (streamwise-shift-reflection and spanwise-shift-rotation) symmetry. 
Equilibria are numerically obtained and overviewed
in \S\ref{sec:overview}, focusing on lower-branch solutions. The dependence of the lower-
and upper-branch equilibria on the computational domain and on the LES parameters are
investigated in \S\ref{sec:tall-EQ}, while \S\ref{sec:upper-branch} and
\S\ref{sec:local-turb} discuss the vertical localisation of the equilibrium solutions and of
the baseline LES. Finally, \S\ref{sec:conc} offers conclusions and future perspectives. An
appendix discusses the linear stability of the equilibrium solutions.

\section{Numerical methods}\la{sec:numer}

\subsection{The governing equations}\label{sec2:LES}

We solve the incompressible LES momentum and continuity equations, 
\begin{eqnarray}\label{eq:LES}
  \partial_t \ubar_i + \ubar_j \partial_j \ubar_i &=&
  - \partial_i \pbar + \partial_j \left( 2 \nu_t \sbar_{ij} \right),\\
    \partial_j\ubar_j &=& 0, \label{eq:LESc}
\end{eqnarray}
where $\ubar_i$ are the resolved (large scale) velocities, and $\pbar$ is a modified
resolved kinematic pressure that includes the diagonal part of the SG stress tensor. The
eddy viscosity is written as $\nu_t = l_S^2 |\sbar|$, in terms of a static length scale
$l_S$ and of the local resolved strain rate \citep{Smag:63}, where $|\sbar|^2=2 \sbar_{ij}
\sbar_{ij}$ and $\sbar_{ij}=(\partial_{j} \ubar_{i} + \partial_{i} \ubar_{j})/2$ is the
strain-rate tensor of the filtered velocity. We use $(x,y,z)$ to represent the streamwise,
vertical (cross-shear) and spanwise coordinates, respectively, and $(u,v,w)$ to denote the
respective velocity components. Whenever convenient, as in (\ref{eq:LES}), this notation is
substituted by subindices, in which case repeated indices imply summation. This is always
the case for the vorticity components $\ombar_i$. There is no explicit filtering in our
code, and the smallest flow scales are controlled either by the grid or by the eddy
viscosity.

It is usual in most LESes to include the molecular viscosity $\nu$ as part of the right-hand
side of (\ref{eq:LES}), and to write the length $l_S$ in terms of some convenient grid spacing
$\Delta_g$ and of a `universal' Smagorinsky constant $C_S$. None of those devices are needed
here, and will not be used. The main role of the molecular viscosity in LES is to enforce
the boundary conditions when $\nu_t$  vanishes at the wall. Since there are no
walls in our problem, we set $\nu=0$, and all our simulations run at infinite molecular
Reynolds number.

Similarly, the role of $\Delta_g$ in SG models is to accommodate variable grid
spacings, assuming that most of the spectral content of the velocity gradients is
concentrated at the grid scale. Again, the statistical homogeneity of the flow implies that
our grids are uniform, and makes this complication unnecessary.

In fact, the essential role of the eddy viscosity is to introduce a minimum length scale
that takes the role of the Kolmogorov length, $\eta$, independently of the grid. If we
estimate the energy dissipation by the subgrid model as
$\dis=\bra\nu_t|\sbar|^2\ket=l_S^2\bra |\sbar|^3\ket$, and define an effective Kolmogorov
length as $\eta_t = (\bra\nu_t\ket^3/\dis)^{1/4}$, it follows from the definition of $\nu_t$
that $\eta_t\approx l_S$. Here $\bra \cdot \ket$ represents time and space average, and will
occasionally be replaced by $\bra \cdot \ket_{xz}$ to represent the $y$-dependent average
over horizontal planes, or by $\bra \cdot \ket_c$ to represent $\bra \cdot \ket_{xz}$
particularised to the central plane $y=0$. From now on, capital letters are reserved
for mean quantities and lower-case ones for fluctuations with respect to that mean, as in
$U=\bra \ubar \ket_{xz}$ and $\ubar= U + u$. Primes denote root-mean-squared (rms)
intensities of the fluctuations.

Since most of the computational effort in DNSes is dedicated to
resolving the smallest dissipative scales, the introduction of a cut-off length in LES
drastically reduces the size of the linear systems that have to be solved to obtain
invariant solutions.

For the parallel flows which are the subject of this paper, (\ref{eq:LES}) can be integrated to
\beq
\bra -uv +2 \nu_t \sbar_{xy}\ket_{xz}=u_\tau^2, 
\la{eq:utau}
\eeq
where $u_\tau$ is independent of $y$ and can be used to scale the velocities. Variables in
this scaling are denoted by a `+' superscript.

\begin{table}
  \centering
  \begin{tabular}{c*{10}{c}}%
    Run  & $A_{xz}$ & $A_{yz}$ & $R_S$ & $Re_z$ & $Re_\lambda $
    & $ L_0/L_z$ &  $\eta/L_z $  & $ u^\prime/SL_z$ & $v^\prime/SL_z$ & $w^\prime/SL_z$  \\
    L32 (DNS) & 3 & 2 & 104 & 2000  & 47 & 0.41 & 0.00965 & 0.208 & 0.159 & 0.164 \\
    M32 (DNS) & 3 & 2 & 367 & 12500 & 105 & 0.38 & 0.00273 & 0.182 & 0.131 & 0.136 \\
   LES$_\mathrm{s}$  & 3 & 1.33 & 50.5 & 1068  & 31.6 & 0.296 & 0.0170  & 0.196 & 0.0915 & 0.118 \\  
   LES$_\mathrm{m}$  & 3 & 1.33 & 91.2 & 1837  & 52.3 & 0.384 & 0.00974  & 0.232 & 0.159 & 0.172 \\  
    LES$_\mathrm{t1}$  & 3 & 3 & 50.0 & 1187  & 42.8 & 0.476  & 0.0173  & 0.225 & 0.0737  & 0.0957 \\  
    LES$_\mathrm{t2}$  & 3 & 3 & 101.6 & 2324  & 55.9 & 0.381  & 0.00875  & 0.212 & 0.141 & 0.158 \\

    LES$_\mathrm{t3}$  & 3 & 3 & 203 & 6577 & 85.5 & 0.354 & 0.00436  & 0.181 & 0.127 & 0.140
  \end{tabular}
\caption{Parameters for the turbulence runs. L32, M32 are reference DNSes of SS-HST from
\cite{SekimotoDongJimenez2016}. LES$_\mathrm{s,m,t1,t2,t3}$ are the present LESes in the
symmetric subspace defined by (I)+(II) in (\ref{eq:sym1})--(\ref{eq:sym2}). The effective Reynolds
number, $Re_z$, and the Kolmogorov viscous scale, $\eta$, are computed with the molecular
viscosity in DNS, and with the averaged eddy viscosity in LES. The scale ratio $R_S$ is
$L_z/l_S$ in LES and $L_z/\eta$ in DNS. $Re_\lambda \equiv u_0 \lambda / \bra \nu_t \ket$ is
the Reynolds number based on the Taylor-microscale $\lambda = \sqrt{15}
u_0/\bra \omega_i \omega_i \ket^{1/2} $.
}
    \label{table:LES-DNS}
\end{table}

\begin{figure}
  \centering
  \includegraphics[width=0.8\linewidth,clip]{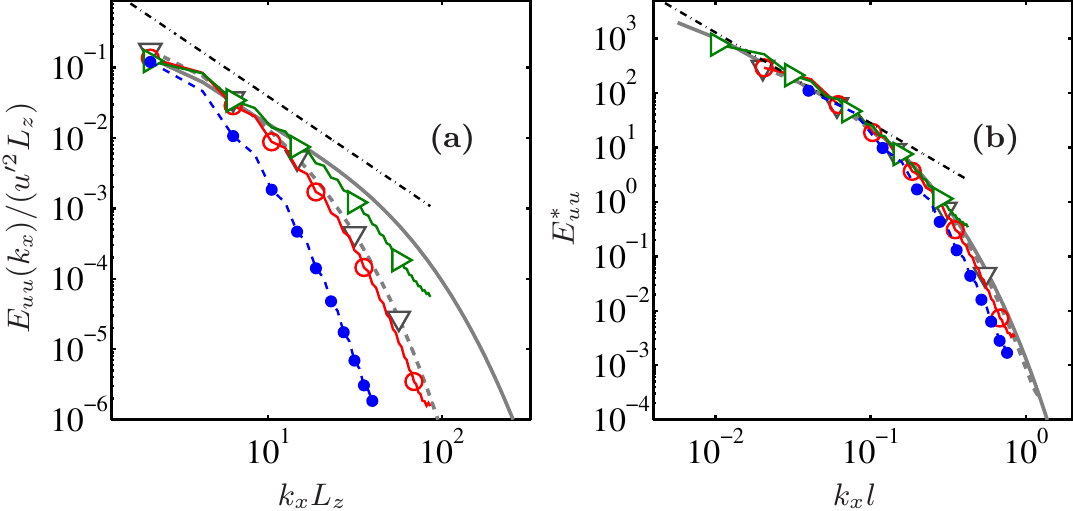}
  \caption{Streamwise velocity spectra $E_{uu}(k_x)$:
    (a) Large-scale normalisation, $E_{uu}/ ( u^{\prime2}L_z)$ as a function of $k_x L_z$.
    (b) Small-scale normalisation, $E_{uu}^\ast \equiv \dis^{-2/3} l^{-5/3} E_{uu}(k_x l)$, with
    $l = l_S$ for LES and $l = \eta$ for DNS.
    \dashedtridown (grey), DNS (L32); 
    \solid (grey), DNS (M32);  
    \dashedcircf (blue), $R_S=52.6$;
    \solidcirc (red), $R_S = 101.6$;
    \solidtriright (green), $R_S = 203$.
    The slope of the chain-dotted diagonal is $-5/3$.
  }
  \label{fig:Spec_LES_DNS} 
\end{figure}

\subsection{Large-eddy simulations with sinuous symmetry}\label{sec2:LES2}

We study a flow with a nominally uniform mean shear $S=\dr U /\dr y$, which is a given constant, 
in a parallelepipedic
computational domain that is periodic in the $(x,z)$ directions, and periodic between points
of the lower and upper boundaries that are uniformly shifted in time by the shear
\citep{GerzSchumannElghobashi1988}.  
The mean flow for turbulence statistics is 
$Sy$ plus the vertically periodic correction, 
which is an approximately linear profile as evaluated later in this section.
The numerical code for DNS is described in detail in
\citet{SekimotoDongJimenez2016}. 
The equations (\ref{eq:LES})--(\ref{eq:LESc}) for LES are formulated in
terms of the vertical vorticity and of the Laplacian of the vertical
velocity~\citep{KimMoinMoser1987,HughesOberaiMazzei2001}. The discretisation uses
2/3-dealiased Fourier expansions in $(x, z)$, and sixth-order spectral-like compact finite
differences in $y$, with the shear-periodic boundary conditions embedded in the compact
finite-difference matrices for each Fourier mode. As explained in
\citet{SekimotoDongJimenez2016}, this avoids recurrent remeshing and the resulting secular
loss of enstrophy over long integration times. 

There are three dimensionless parameters: a Reynolds number, and two box aspect ratios,
$A_{xz} = L_x/L_z$ and $A_{yz} = L_y/L_z$, where the $L_j$ are the dimensions of the
computational domain along the three coordinate directions. It was shown in
\citet{SekimotoDongJimenez2016} that the correct scales for the large-scale length and velocity of
SS-HST are based on the spanwise box dimension, $L_z$ and $SL_z$, so that the 
Reynolds number is $Re_z \equiv SL_z^2/\nu$ in DNS. We will extend this definition to LES using
the effective mean eddy viscosity, $\bra\nu_t\ket$, replacing the molecular viscosity $\nu$.  
However, it is more convenient to use the length-scale ratio $R_S = L_z/l_S$ in LES. 
For example, \cite{PiomelliRouhiGeurts2015} recently developed a grid-independent LES using as 
parameter the ratio of the small (effective Kolmogorov) and integral scales, which is
essentially the inverse of $R_S$.
It can be extended to DNS by substituting $\eta$ for $l_S$ ($L_z/\eta \sim Re_z^{3/4}$) and  
it turns out that, in DNS of SS-HST, $Re_z \approx 4 R_S^{4/3}$ \citep{SekimotoDongJimenez2016}.
We will also use the integral scale $L_0= u_0^3/\dis$,  where $u_0^2=\bra u_i u_i \ket/3$.

We study in this paper LES turbulence and invariant solutions in a subspace defined by
enforcing two `sinuous' symmetries: a reflection with respect to the plane of $z=0$ followed by a
streamwise shift by $L_x/2$,
\beq\la{eq:sym1}
(\mathrm{I}):~[u,v,w](x,y,z) = [u,v,-w](x+L_x/2,y,-z), 
\eeq
and a rotation by $\pi$ around $x=y=0$ followed by a spanwise shift by $L_z/2$,
\beq\la{eq:sym2}
(\mathrm{II}):~[u,v,w](x,y,z) = [-u,-v,w](-x,-y,z+L_z/2).
\eeq
Note that no translational symmetries are allowed in this subspace, so that travelling waves
are excluded. Moreover, $(\mathrm{I})$--$(\mathrm{II})$ together with the boundary conditions
enforce that the instantaneous plane-averaged streamwise velocities
at the top and bottom of the box are $U (\pm L_y/2) =\pm S L_y/2$.

Before proceeding to seek equilibrium solutions, the effect of the symmetry restriction was
tested in several LESes of symmetric SS-HST, which are summarised in
table~\ref{table:LES-DNS}. The table also includes two reference unconstrained DNSes from
\citet{SekimotoDongJimenez2016}. Table~\ref{table:LES-DNS} shows that the length and
velocity scales found in DNS also work well in the symmetric LESes. Although not included in
the table, the effective Kolmogorov scale in the LESes is found to be $\eta_t/l_S \approx
0.9$, in approximate agreement with our analysis in \S\ref{sec2:LES}. Interestingly, both
$\eta_t/l_S \approx 0.9$--$1.0$ and $L_0/L_z \approx 0.4$--$0.7$ remain good approximations
in the localised equilibrium solutions described below. Note that the vertical translational
invariance is broken by the symmetries (\ref{eq:sym1})--(\ref{eq:sym2}), so that a uniform mean
shear is no longer guaranteed. Even so, the LESes of symmetric turbulence retain an
approximately linear mean profile $(|U-Sy|\approx 0.01SL_z)$ and approximately constant
fluctuation profiles. We will see below that the same is not generally true for the profiles
of the equilibrium solutions.
 
Figure~\ref{fig:Spec_LES_DNS} displays the longitudinal streamwise-velocity spectrum of the
symmetric LES turbulence, compared to that of the reference DNSes. Given the different
Reynolds numbers and techniques, the agreement is good, showing that symmetry does not
influence turbulence greatly. Note that the three LESes in figure~\ref{fig:Spec_LES_DNS}(b)
collapse well in terms of $l_S$, and that the resolution of the simulations is fine enough to
resolve the smallest LES scales.
  
The size of the collocation grids for the turbulence LESes and for typical equilibrium
solutions are given in table~\ref{table:EQ-grid} of \S\ref{sec:overview}. It follows from
the definition of the different quantities that the collocation resolution is 
$\Delta x/l_S = R_S A_{xz}/N_x$, $\Delta y/l_S = R_S A_{yz}/N_y$ and $\Delta z/l_S = R_S/N_z$,
where the $N_j$ are the collocation points along the three coordinate directions.
For example, in the lowest-Reynolds number LES$_{\rm s}$ in tables
\ref{fig:Spec_LES_DNS} and \ref{table:EQ-grid}, this results into
$(\Delta x,\Delta y,\Delta z)=(3.6, 1.4, 2.7) l_S$. If our SG model is interpreted in terms
of the usual Smagorinsky formula $\nu_t=(C_S \Delta)^2 |\sbar|$, with $\Delta$ a
representative grid spacing, it follows that $C_S=l_S/\Delta = 0.42$, which is a
relatively high value for usual LES practice. In this sense, all our LESes are overdamped.
Their dynamics are not controlled by the grid, but by the SG model, and can be expected to
be approximately independent of the resolution. They also only resolved the integral scales
of the flow, and very little of its inertial range. The effective Smagorinsky constant for
the different simulations is included in table \ref{table:EQ-grid}.

At box aspect ratios $(A_{xz},A_{yz})=(3,1.33)$, symmetric self-sustaining LES turbulence
exists for $R_S \gtrsim 65$. There is a transitional range, $45 \lesssim R_S \lesssim 65$,
in which the kinetic energy of the flow increases and decreases several times before
decaying to laminar after $St=O(1000)$. Both the transitional range and the decay time
depend somewhat on $A_{yz}$. We will see in \S 4 that LES turbulence in taller boxes, $A_{yz}\approx 2$--$3$,
survives in the range $50 \lesssim R_S \lesssim 60$, but collapses intermittently to
vertically localised turbulence.

\subsection{Searching for invariant solutions in LES}\label{sec2-2:howtosearch}

Our main interest is to characterise invariant solutions in LES of SS-HST.
Strictly equilibrium solutions are technically impossible in this system.
The shear-periodic boundary condition slides the upper computational boundary with a velocity
$SL_y$ with respect to the lower one, and the numerical configuration only repeats itself
after a `box' period $T_s \equiv L_x/(SL_y)$. It was shown in
\citet{SekimotoDongJimenez2016} that this periodic forcing does not interfere strongly with
the turbulent solutions as long as the aspect ratios are kept in the range $2< A_{xz}
\lesssim 5$ and $A_{xz} \lesssim 2 A_{yz}$. We will approximately respect these constraints
here, and find solutions that are numerically indistinguishable from fixed points. 
Vertically localised solutions can be equilibria at the limit of $A_{yz} \rightarrow \infty$, since they are independent of the shear-periodic boundary condition, but it should be remembered that all of them are conceptually periodic orbits 
in a finite computational domain. 
All of the statistics discussed below are averages over a box period.

Solutions are computed using the Newton--Krylov--hookstep method~\citep{Viswanath2007} on
$\fvec^{T} (\xvec) - \xvec = 0 $, where $\xvec$ is the vector of independent variables,
and $\fvec^{T}: \xvec(0) \rightarrow \xvec(T)$ is the integration of (\ref{eq:LES}) over time
$T$, using the evolution code described in the previous section. Because of the periodicity
mentioned above, the search time is always an integer multiple of the box period, 
$T_m = m T_s$. The convergence criterion for the relative error of the Newton method is generally taken
$\|\fvec^{T_m}(\xvec) - \xvec\|/\|\xvec\|< 10^{-5}$, where $\|\cdot\|$ is the L$_2$ norm.

The initial condition for the search was taken from snapshots of the LES symmetric
turbulence in the above-mentioned marginal range in which turbulence eventually decays to
laminar, $R_S <50$, $A_{xz}=3$ and $A_{yz}=1.33$. 
In order to find a solution, we fix the aspect ratios and the time period $T_m \approx 10 $ ($m=5$), then change $R_S$.
For $R_S \approx 35$, the Newton search always converges to a trivial laminar state. 
The first non-trivial solution is obtained spontaneously at $R_S = 38.6$, 
and a trial with $m=7$ converges to the same solution. 
It is, therefore, a periodic orbit with the period of $T_m = T_s$ ($m=1$) close to the bifurcation point that marks the lowest $R_S$ of our family of solutions. 
The solutions are always unstable and some example of the linear stability analysis are shown in the later section and in the Appendix~\ref{appdx:floquet}.

From this initial solution, lower and upper branches are continued along the parameters
$A_{xz},~A_{yz}$ and $R_S$, overcoming the turning points with a pseudo-arclength method. We
have seen that all these solutions are periodic orbits in which the period is $T_s$. The periodicity is especially noticeable for the upper branch solutions in flat
boxes, $A_{yz}=1.33$--$1.64$ and $R_S \approx 45$, where the temporal oscillation of the
integrated kinetic energy is of order 1\%. 
They asymptotically approach a fixed point in the limit of $A_{yz} \rightarrow \infty$, and the oscillations become negligible, $O(10^{-5})$ for $A_{yz}>2$, 
for the vertically localised solutions described later in taller boxes.

\begin{figure}
  \centering
  \includegraphics[width=0.8\linewidth,clip]{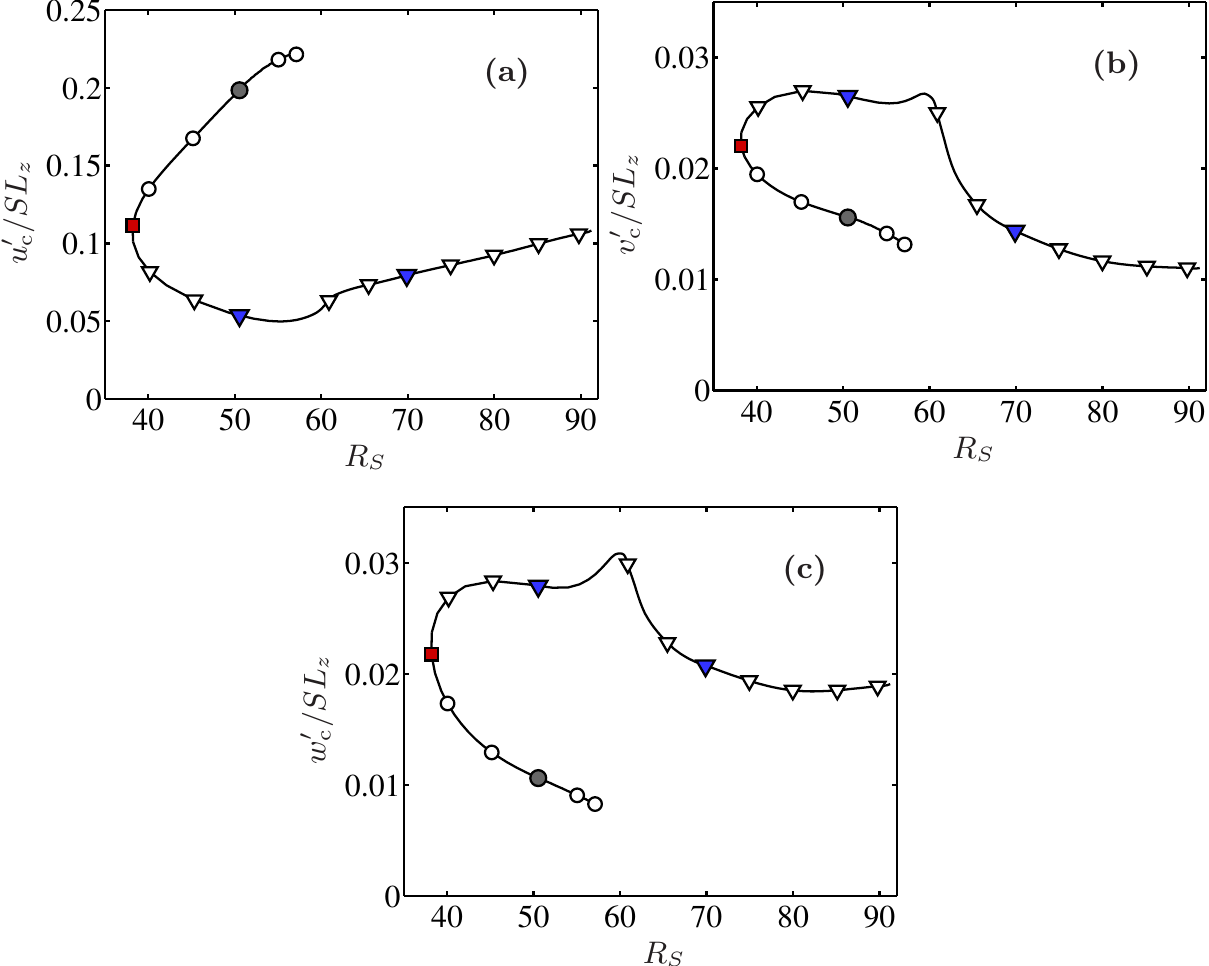}
\caption{(a-c) Velocity fluctuation intensities at the centre plane $y=0$, as functions of
$R_S=L_z/l_S$, for $(A_{xz},A_{yz})=(3,1.33)$.
(a) Streamwise component, $u^\prime_c$. (b) Vertical, $v^\prime_c$. (c) Spanwise
$w^\prime_c$.
($\circ$, black) the upper branch; ($\triangledown$, blue) the lower branch. Solid symbols
are cases plotted in figure \ref{fig:UPO_LES_flat_3d}.
  }
\label{fig:bifurcation} 
\end{figure}
\begin{figure}
  \centering
  \includegraphics[width=0.7\linewidth,clip]{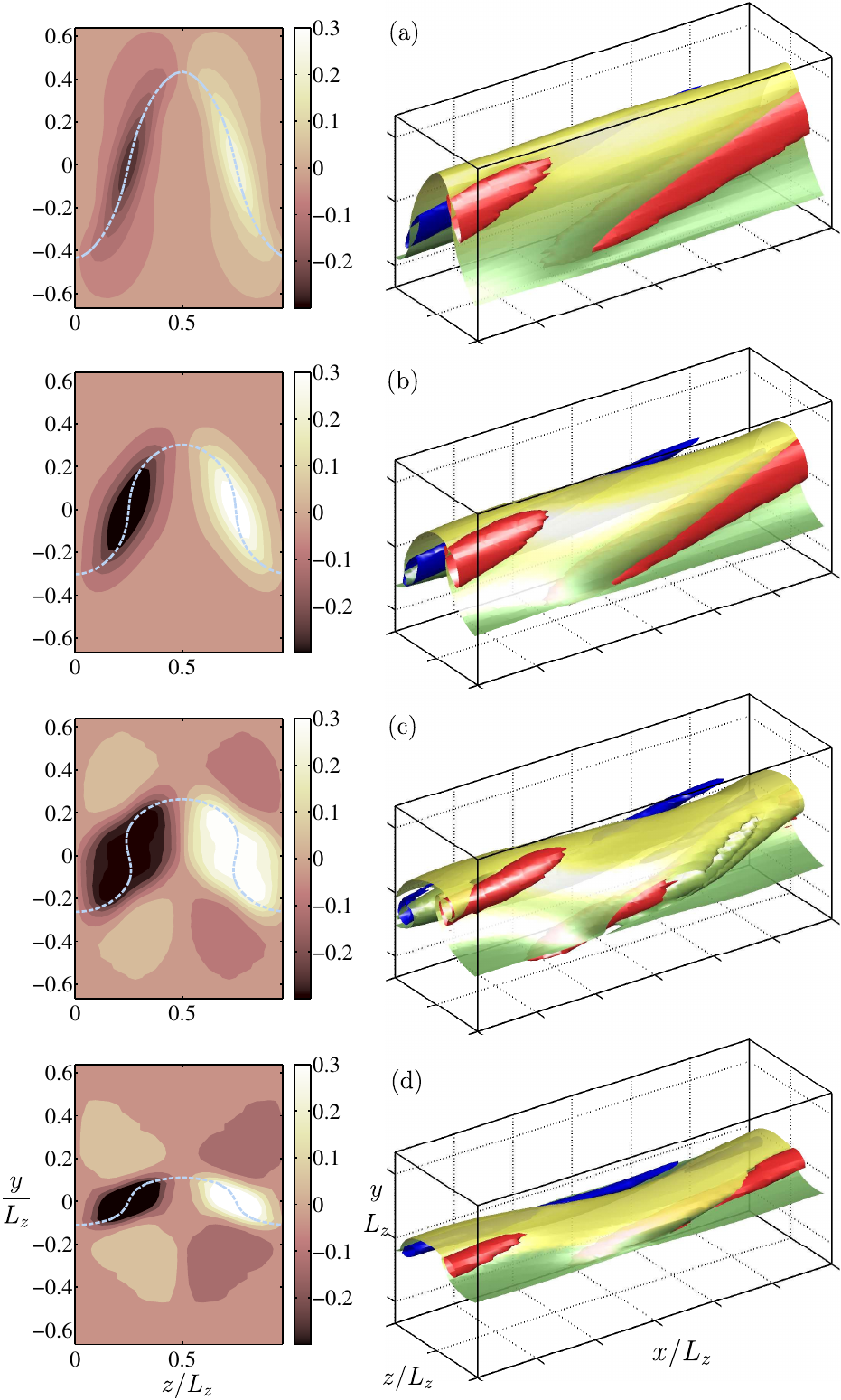}
  \caption{Vortical structures and the velocity streaks in a box $(A_{xz},A_{yz})=(3,1.33)$: 
(a) $R_S = 50.5$, upper branch. 
(b) $R_S =  38.2 $ at the bifurcation point. 
(c) $R_S =  50.5$, lower branch. 
(d) $R_S =  69.8$, lower branch.
The shaded background in the left figures are isocontours the streamwise-averaged
$\omega_x(y,z)/S =[-0.3:0.03:0.3]$. The light-grey dashed line is the streamwise-averaged
$\ubar=0$.
The right figures shows the isosurfaces of:
Red (dark-grey in black/white), $\omega_x = 0.6 |\omega_x|_\mathrm{max}$; 
blue (black), $\omega_x =- 0.6 |\omega_x|_\mathrm{max}$; 
yellow (light-grey), $\ubar=0$ gradually coloured by $y$. 
  }
  \label{fig:UPO_LES_flat_3d} 
\end{figure}
\begin{figure}
  \centering
  \includegraphics[width=0.8\linewidth,clip]{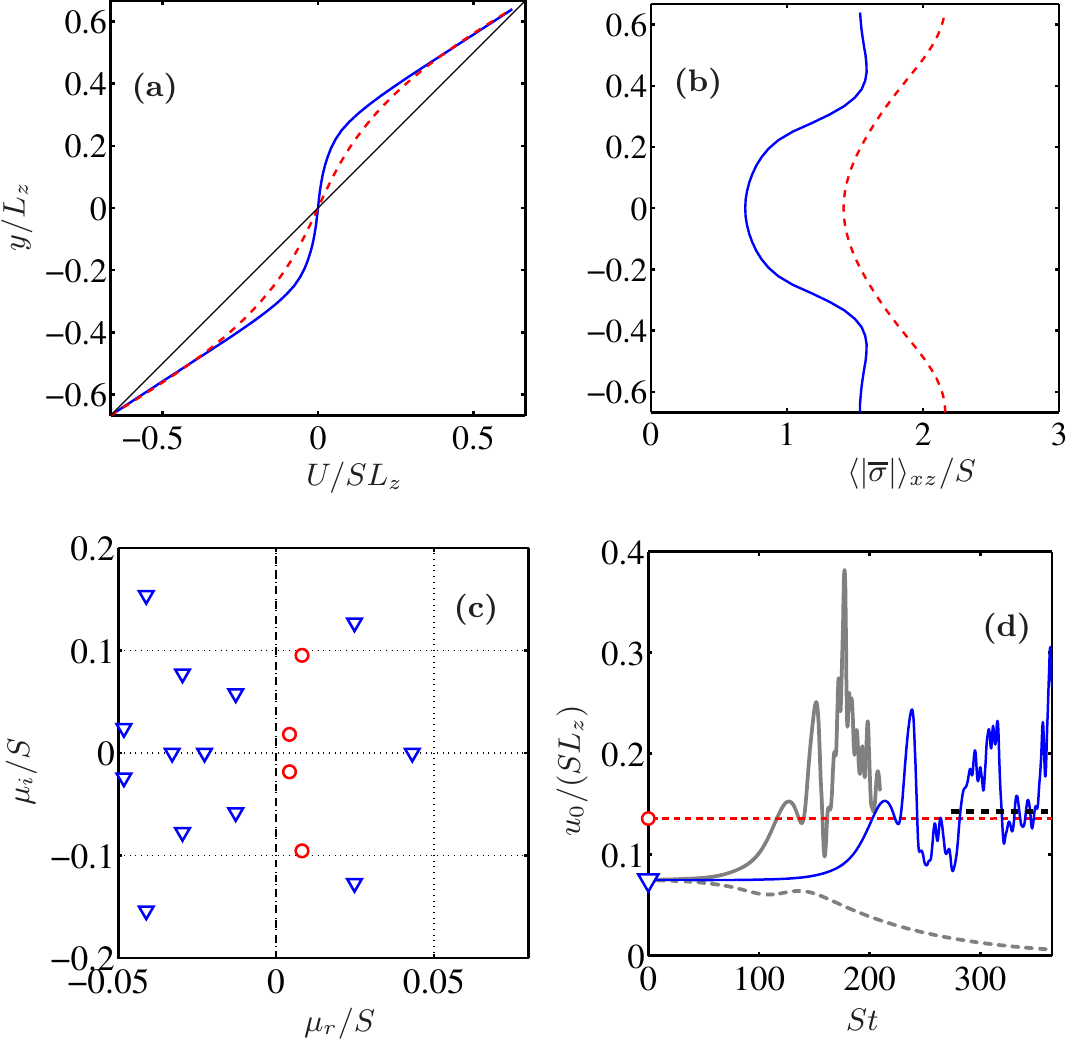}
\caption{LES equilibrium solutions, $(A_{xz},A_{yz})=(3,1.33)$ and $R_S=50.5$. 
In (a,b): \solid, Lower branch, as in figure~\ref{fig:UPO_LES_flat_3d}(a); 
\dashed, upper branch, as in figure~\ref{fig:UPO_LES_flat_3d}(c).
(a) Mean streamwise velocity. The  thin diagonal is $U = Sy$.
(b) Resolved strain rate. 
(c) Stability eigenvalues $(\mu_r + \ii \mu_i)/S$ of the equilibria in (a,b): 
$\triangledown$ (blue), lower branch; $\circ$ (red), upper branch. 
(d)  Temporal evolution of the fluctuation velocity magnitude of symmetric LES initialised from the equilibria in (a,b). 
The darker lines are initialised without any disturbance except numerical inaccuracies. 
\solid (dark blue), Initialised from the lower branch;
\dashed (dark red), from the upper branch. 
The lighter grey lines are initialised from the lower-branch with a small disturbance along
the unstable direction corresponding to the real unstable mode. 
\solid, attracted to the turbulence state; 
\dashed, attracted to the laminar state. 
All attempts to perturb the upper branch along its unstable modes led to laminarisation. The
short thick dotted line represents the fluctuation intensity of LES$_\mathrm{s}$ in
table~\ref{table:LES-DNS}, using the same box and Reynolds number.
}
 \label{fig:UPO_LES_2} 
\end{figure}

\section{Lower-branch solutions in LES}\label{sec:overview}

Here, we preview the obtained equilibria, and characterise lower-branch solutions
in a computational box, $(A_{xz}, A_{yz})=(3, 1.33)$. This is slightly `flatter' than the
acceptable range, $A_{xz} \lesssim 2 A_{yz}$, in which unphysical strong linear bursts appear as described in \citet{SekimotoDongJimenez2016},
in spite of which the turbulence statistics are reasonably close to the logarithmic layer of
channel turbulence. We investigate in this section the $R_S$ dependence of the equilibria.
The effect of the aspect ratios is discussed in \S\ref{sec:box-dependency}.

\begin{figure}
  \centering
  \includegraphics[width=0.8\linewidth,clip]{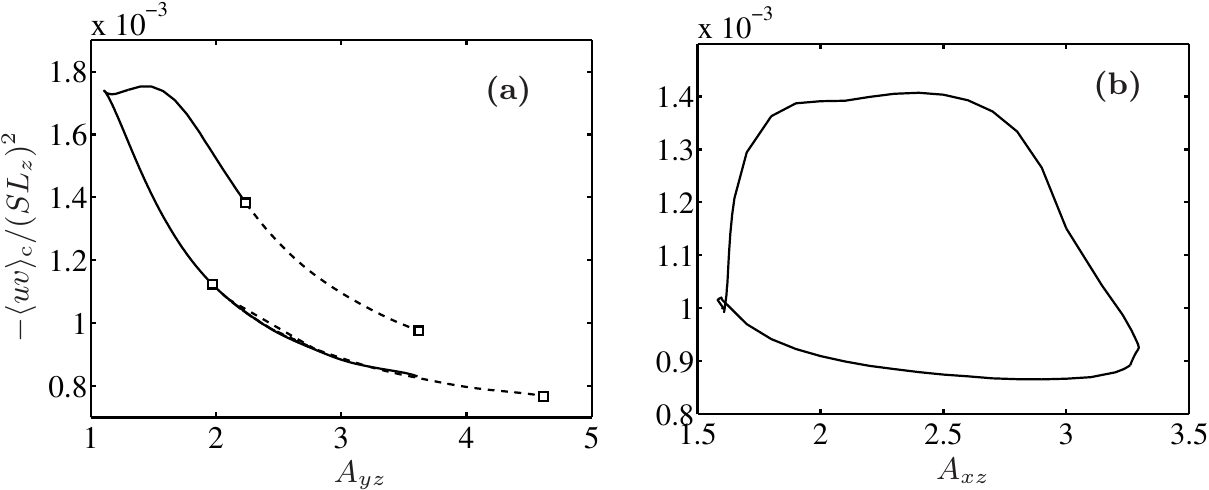}
  \caption{Reynolds stress averaged over the $y=0$  plane, $R_S=38.9$, $N_x=64$ and $N_z=32$.
(a) As a function of  $A_{yz}$ for $A_{xz}=3$.  \solid, $N_y=48$; $\square$\dashed$\square$, $N_y=48$. 
(b) As in (a),  a function of $A_{xz}$ for $A_{yz}=3$, $N_y = 64$.
  }
  \label{fig:LES_bif_box_dependency} 
\end{figure}

All the equilibria in this section are found using the same numerical grid of the low-$R_S$
turbulence LES$_{\rm s}$ in table \ref{table:EQ-grid}, $(N_x,N_y,N_z)=(64,48,32)$. They are
continued in $R_S$ as long as convergence is achieved. This happens in $R_S \lesssim 100$
for the lower-branch solutions, and in $R_S \lesssim 60$ for the upper branch. We emphasise
that $R_S$ depends on the eddy-viscosity parameter $l_S$, rather than on the numerical grid.
As long as the numerics resolves features of the order of $\eta_t\approx l_S$, all grids
should converge to the same solution. We saw in the previous section that this is true for
most of our simulations. As a further check, finer grids are used in \S\ref{sec:tall-EQ} to
compute equilibria in other numerical boxes. In the cases in which the same solution is
computed with two different grids, no substantial difference are found (see figure
\ref{fig:LES_bif_box_dependency}).

Although most solutions were converged to a Newton relative error of $ 10^{-5}$, a few were
confirmed to a tighter tolerance. For example, the lower- and upper-branch solutions at
$R_S=50.5$ discussed in figures \ref{fig:UPO_LES_flat_3d} and\ref{fig:UPO_LES_2} are iterated to
an accuracy of $10^{-12}$.

Figures~\ref{fig:bifurcation}(a--c) are continuation diagrams of the lower- and upper-branch
solutions for the box mentioned above, showing the fluctuation intensity of the three
velocity components at $y=0$. The lower-branch solutions are characterised by weaker
streamwise velocity fluctuations and stronger transverse velocities than those in
the upper branch, but note that the even those more intense transverse velocity
fluctuations, $v^\prime$ and $w^\prime$, are an order of magnitude smaller than in the LES and
DNS turbulence at a comparable Reynolds number in table~\ref{table:LES-DNS} $(v'\approx w'
\approx 0.16SL_z)$.

\begin{figure}
  \centering
  \includegraphics[width=0.8\linewidth,clip]{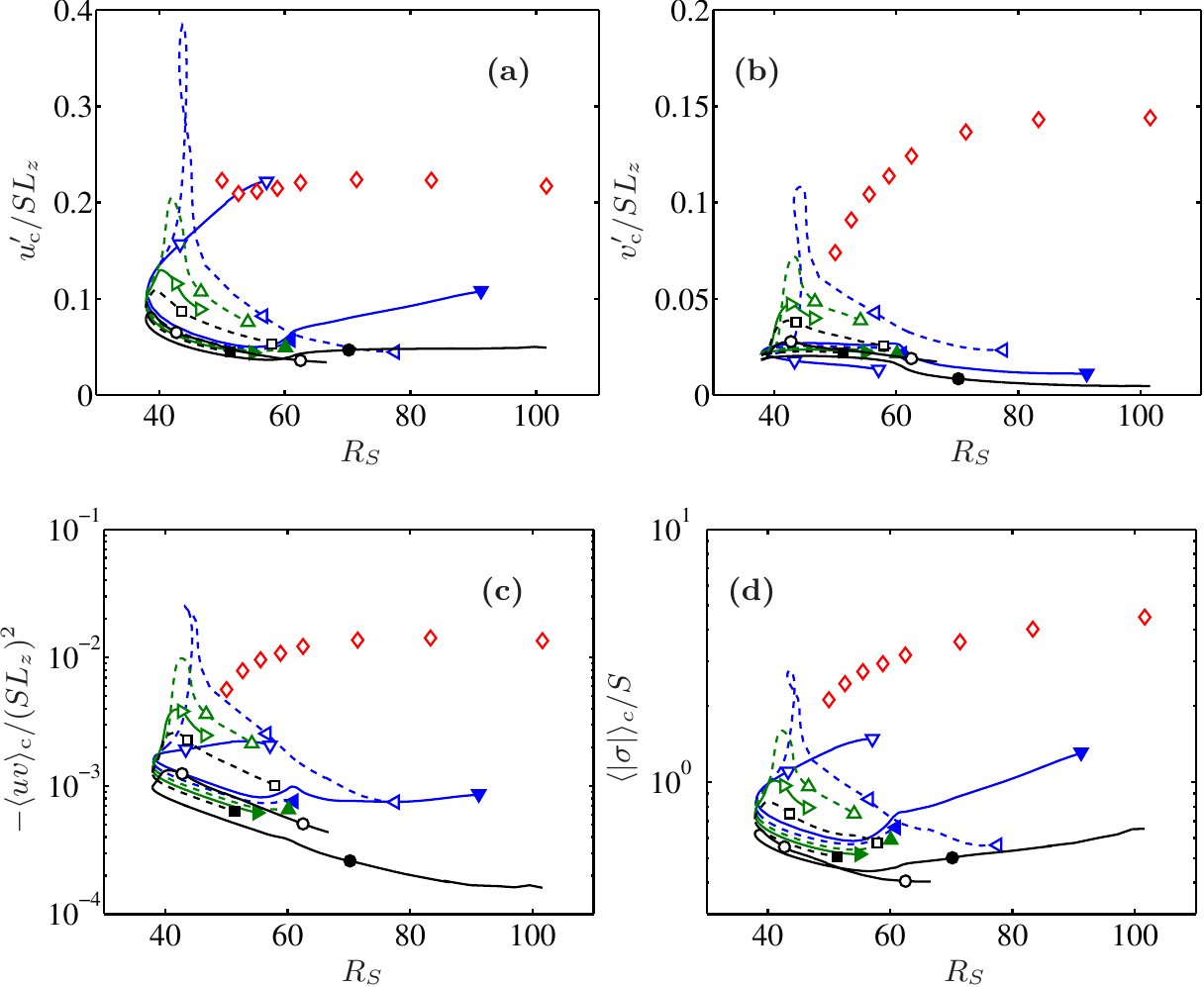}
 \caption{Equilibrium solutions for $A_{xz}=3$ and different $A_{yz}$:
    (solid with $\triangledown$, blue ) $A_{yz}=1.33$; 
    (dashed with $\triangleleft$, blue) $1.5$; 
    (dashed with $\triangle$, green) $1.64$; 
    (solid with $\triangleright$, green) $1.8$; 
    (dashed with $\square$, black) $2.0$; 
    (solid with $\circ$, black) $3.0$. 
    (a) $u^\prime_\mathrm{c}/SL_z$. (b) $v^\prime_\mathrm{c}/SL_z$. 
    (c) Resolved-scale Reynolds stress $-\langle u v \rangle_c$.
    (d) $\langle |\sigma| \rangle_{c} /S \equiv R_S^2/\langle Re_z \rangle_c$.
    The filled and open symbols represent lower and upper branches, respectively. 
    The red diamonds are box-averaged statistics of  symmetric LES turbulence in a box $(A_{xz},A_{yz})=(3,3)$.
    }
 \label{fig:LES_bif_Ax3_invCz} 
\end{figure}

The identification of which branch is the lower one is not straightforward, but we will denote 
as such the lower branch in figure \ref{fig:bifurcation}(a). 
This is actually consistent with the fact that the lower-branch solution typically has low drag, while the upper one has high drag (figure \ref{fig:LES_bif_Ax3_invCz} c) as we shall discuss later in \S\ref{sec:box-dependency}.
Continuation in $R_S$ reveals
that, as this branch extends towards higher $R_S$ its solutions concentrate in a relatively
thin critical layer for $R_S > 70$. This is a common feature of lower-branch solutions in
wall-bounded flows at high Reynolds numbers
\citep{WangGibsonWaleffe2007,Viswanath2009,HallSherwin2010,%
DeguchiHallWalton2013,BlackburnHallSherwin2013}.
Figure~\ref{fig:UPO_LES_flat_3d} shows isosurfaces of $|\omega_x| =
0.6|\omega_x|_\mathrm{max}$, and of $\ubar = 0$, representing the geometry of the
vortical structures and of the velocity streak in the upper- and lower-branch solutions,
showing that they are localised around $y=0$. 
The streamwise-velocity streak of the lower-branch solutions meanders more deeply than the one in
the upper-branch, leading to stronger cross-flow velocity fluctuations. As $R_S$ increases,
the vortical structure of lower-branch solutions becomes flatter. This occurs drastically,
but smoothly, at around $R_S \approx 60$, after which it becomes a critical-layer-type
solution similar to those described by the vortex-wave interaction (VWI) theory for the
lower-branch solutions in plane Couette flow \citep{BlackburnHallSherwin2013}. 
This transition probably corresponds to a 
complex set of bifurcations, since several eigenvalues change stability around that $R_S$,
but tracking them is difficult in our flow. For example, a new complex pair appears
around $R_S=55$, but the associated branch cannot be followed by the present method because
its period is not a simple multiple of the box period $T_s$.

In the high-Reynolds number limit, \cite{BlackburnHallSherwin2013} have shown that the VWI
states begin to localise at $y=0$ as spanwise wavenumbers increasing, i.e. as $L_z$ narrows.
\cite{Waleffe1997} showed that equilibrium solutions similar to those of Couette flow are generic to many shear flows, and \cite{DeguchiHall2014_PSTA,Deguchi2015} have described more recently
how a VWI state can be embedded in any shear flow at high-Reynolds number.
There is an inviscid mechanism as in VWI-type solutions of the Navier-Stokes equation, whose streamwise velocity structure is shown to be thinner as increasing $Re$. 
``The singularity occurs where the wave propagates downstream with the local fluid velocity and defines the location of a critical layer in which viscosity smooths out the singularity''~\citep{DeguchiHall2016}. The critical layers in LES equilibria must be similar to those in DNS and the singularity occurs, but the eddy viscosity smooths it out.

Upper-branch solutions are characterised by their taller streamwise-velocity streaks. Their
height increases with increasing $R_S$, while their quasi-streamwise vortices become more
inclined (see figure~\ref{fig:UPO_LES_flat_3d}a). The relatively tall streaks suggest that
the effect of the vertical box aspect ratio may be important for these solutions. This
will be investigated in the next section.

Mean profiles for upper and lower branch solutions at the same $R_S$ are shown in
figures~\ref{fig:UPO_LES_2}(a,b). The lower branch solutions is the one shown in
figure~\ref{fig:UPO_LES_flat_3d}(c), and the upper branch one is in
figure~\ref{fig:UPO_LES_flat_3d}(a). Both solutions are concentrated around $y=0$, but the
concentration is more pronounced in the lower branch. This is seen in the more horizontal
quasi-streamwise rollers in figure \ref{fig:UPO_LES_flat_3d}(c), and in the shallower mean
velocity profile near $y=0$ in figure~\ref{fig:UPO_LES_2}(a). 

Since it follows from the momentum conservation equation (\ref{eq:utau}) that the total shear
stress has to be independent of $y$, the shallower profile of the lower branch suggests a
higher eddy viscosity, and consequently a higher total strain. The opposite turns out to be
the case. Figure~\ref{fig:UPO_LES_2}(b) displays the mean profile of the mean total strain
rate for the two solutions, which can also be interpreted as a profile of eddy viscosity.
The flatter profile of the lower branch is due to higher resolved Reynolds stresses. Beyond
$|y|/L_z \approx 0.5$, the mean strain tends to $\bra|\sbar|\ket_{xz} \approx \dr U/\dr y$,
and most of the momentum flux is carried by the SG term.

The simplest interpretation is that the Reynolds stresses created by the transverse
velocities of the equilibrium state flatten the profile into a local region of lower shear.
This results in a locally lower eddy viscosity and a locally higher Reynolds number, that
helps to sustain the solution. However, the requirement from the boundary condition that the
total velocity difference across the domain is constant prevents the low-shear layer from
spreading over the whole box, and results in a local high-Reynolds number `turbulent' layer
within a box in which all other fluctuations are damped by the model.

The linear stability eigenvalues of the two solutions in figures~\ref{fig:UPO_LES_2}(a,b)
are shown in figure~\ref{fig:UPO_LES_2}(c). The upper-branch has two unstable
complex-conjugate pairs, while the lower-branch solution has a pair of unstable
complex-conjugate modes and a real unstable mode. Since we have already noted that all
solutions are periodic orbits, all these eigenvalues are actually Floquet
exponents that have an underlying periodic component. 
The period of the real unstable mode
of the lower branch in figure~\ref{fig:UPO_LES_2}(c) is the box period, representing an
exponentially growing oscillation synchronous with the numerics. The complex conjugates
pairs have periods that are not simple multiples of the box period, and represent
bifurcations into a torus. Further details of the distribution of the unstable modes and
their dependency on $A_{yz}$ are in the Appendix~\ref{appdx:floquet}.

Figure~\ref{fig:UPO_LES_2}(d) shows the results of
initialising symmetric LESes from the equilibria just discussed. At first, the LES is
initiated from the equilibrium without adding any disturbances beyond numerical errors. The
result is that the lower branch transitions rather quickly to a turbulence-looking bursting state,
while the upper branch does not separate from equilibrium during the time plotted in the figure. 
This is consistent with the stability analysis in figure~\ref{fig:UPO_LES_2}(c), which shows
that the instability eigenvalues of the upper branch are weaker than for the
lower one. 

Next, the LESes are initiated by perturbing the equilibria along the eigenfunction of
individual unstable modes. The grey lines in figure~\ref{fig:UPO_LES_2}(d) show the result
of perturbations of the lower branch along the eigenfunction of its real unstable
eigenvalue. One direction leads to exponential growth of the kinetic energy into a burst and
chaotic turbulence, while the opposite direction laminarises. None of the LESes initialised
from the unstable complex modes of the lower branch leads to bursts or to self-sustaining
turbulence, and neither do the perturbations of the upper branch. The lower-branch solution
thus behaves as a torus `edge', which not only has a single unstable real mode, like simple 
`edge states', but also two complex unstable modes. However, the most
interesting part of this observation is not the detail of this `edge', but the burst
originating from the unstable manifold of the real saddle. A similar behaviour was found by
\cite{VanVeenKawahara2011} in Couette flow.

%
  \begin{table}
    \begin{tabular}{
        c*{7}{c}
      }
      Run & $A_{xz}$ & $A_{yz}$ & $R_S$ & $N_x, N_y, N_z$ & $C_S$ & $\Delta_g/L_z$  \\
      EQ(figure~\ref{fig:UPO_LES_2}) & $3$ & $1.33$ & $50.5$ & $ 64, 48, 32$ & 0.42 & 0.0471  \\
      EQs(figure~\ref{fig:LES_bif_box_dependency}a)  & $3$ & $1.1$--$4.6$ & $38.60$ & $64, 48$--$64$,$32$ & 0.395--0.586 &  0.0442--0.0656 \\
      EQs(figure~\ref{fig:LES_bif_box_dependency}b)  & $1.58$--$3.29$ & $3$ & $38.95$ & $64, 96, 32$ & 0.508--0.649 & 0.0396--0.0505  \\
\\ 
      tall EQ(L)  & $3$ & $3$ & $37.9$--$74.6$ & $ 64, 96, 32$  & 0.274--0.539 & 0.0489 \\
      tall EQ(L)  & $3$ & $3$ & $77.3$--$99.5$ & $ 64, 192, 64$ & 0.334--0.430 & 0.0301 \\
      tall EQ(L)  & $3$ & $3$ & $99.6$--$101.6$& $ 64, 256, 64$ & 0.360--0.368 & 0.0273 \\
      tall EQ(U)  & $3$ & $3$ & $37.9$--$47.4$ & $ 64, 96, 32$  & 0.431--0.538 & 0.0490 \\
      tall EQ(U)  & $3$ & $3$ & $47.4$--$51.5$ & $ 64, 192, 64$ & 0.646--0.702 & 0.0301 \\
      tall EQ(U)  & $3$ & $3$ & $52.4$--$52.6$ & $ 64, 256, 64$ & 0.695--0.698 & 0.0273 \\
      tall EQ(U)  & $3$ & $3$ & $55.5$--$64.5$ & $ 64, 384, 64$ & 0.649--0.754 & 0.0239 \\
\\       
    LES$_\mathrm{s}$ & $3$ & $1.33$ & $50.5$ & $64,48,32$ & 0.42 & 0.0471   \\     
    LES$_\mathrm{m}$ & $3$ & $1.33$ & $91.2$ & $64,64,32$ & 0.256 & 0.0428   \\     
    LES$_\mathrm{t1}$ & $3$ & $3$ & $50.0$ & $64,384,64$ & 0.838 & 0.0259   \\  
    LES$_\mathrm{t2}$ & $3$ & $3$ & $102$ & $128,384,64$ & 0.519 & 0.0190  \\ 
    LES$_\mathrm{t3}$ & $3$ & $3$ & $203$ & $128,384,64$ & 0.260 & 0.0190 \\  
    \end{tabular}
\caption{Grid information for the sinuous-symmetric equilibrium and turbulence LES. EQ(L)
and (U) represent lower- and upper-branch equilibrium solutions, respectively, and the
turbulence LESes are defined in table \ref{table:LES-DNS}. $C_S$ is the effective
Smagorinsky constant discussed in \S\ref{sec2:LES2}, with the filter scale defined as
$\Delta_g \equiv \sqrt[3]{\Delta x \Delta y \Delta z}$.
    }
    \label{table:EQ-grid}
  \end{table}


\begin{figure}
  \centering
  \includegraphics[width=0.8\linewidth,clip]{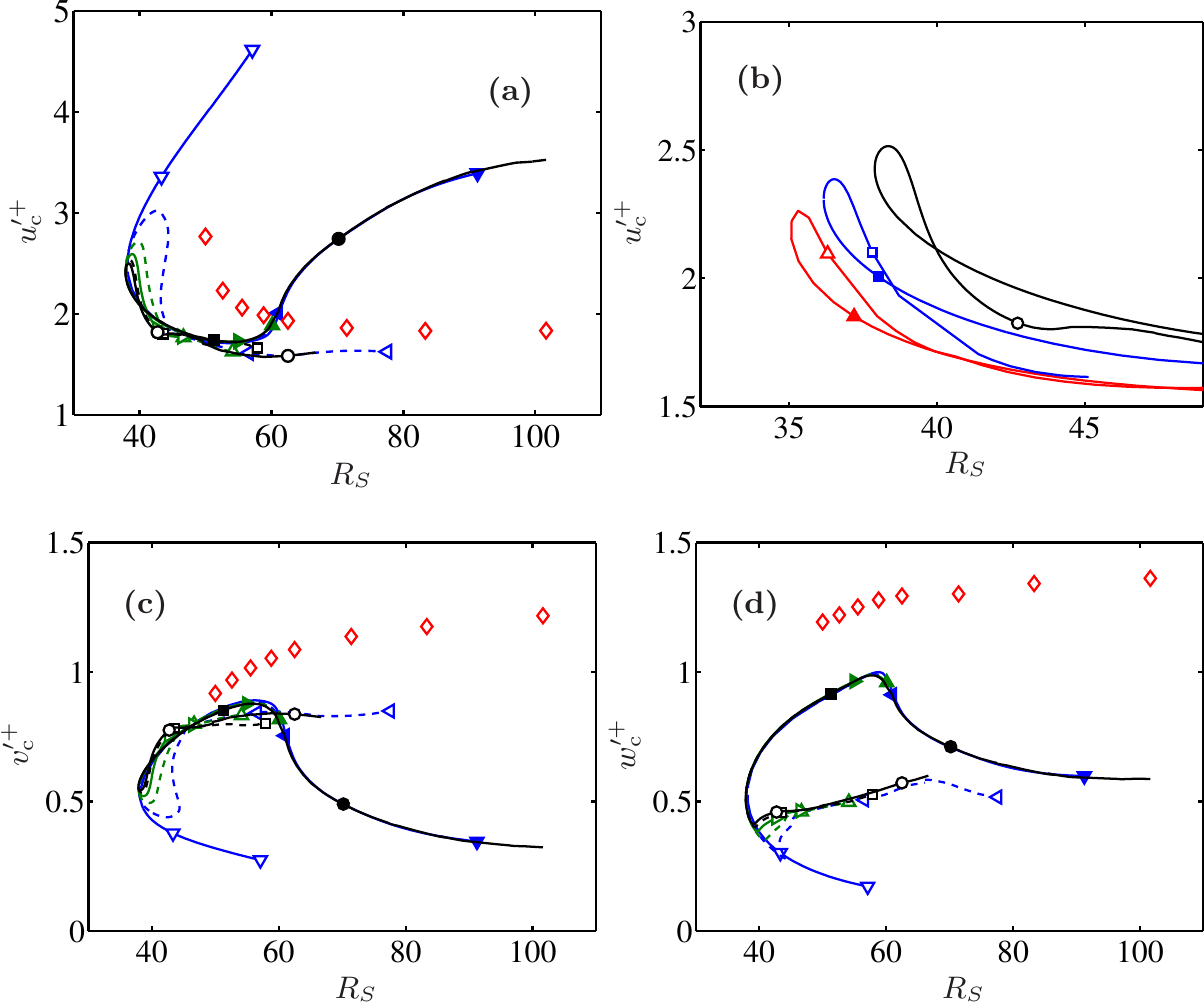}
\caption{ 
    (a,c,d) As in figure~\ref{fig:LES_bif_Ax3_invCz}, but scaled by 
    $u_\tau$ at $y=0$. 
    (a,b) $u^{\prime +}_c$, (c) $v^{\prime +}_c$, (d) $w^{\prime +}_c$. 
    (b) Detail of (a):  
    (solid with $\triangle$, red) $A_{xz} = 2$, 
    (solid with $\square$, blue) $A_{xz} = 2.5$, 
    (black with $\circ$) $A_{xz}=3$.
  }
  \label{fig:LES_bif_Ax3_invCz_wall_unit} 
\end{figure}

\begin{figure}
  \centering
  \includegraphics[width=0.8\linewidth,clip]{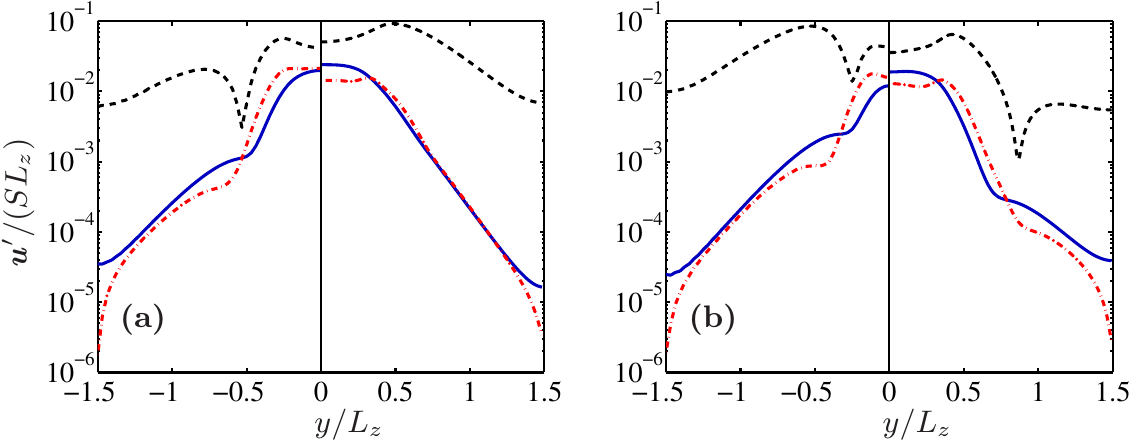}
  \caption{Velocity fluctuations for $A_{xz} = A_{yz} = 3$. (a) $R_S=50$, (b) $R_S=62.5$ 
    Only half of each plot is shown, using the symmetry in y.  Left side is lower branch. Right side is upper branch.
    \dashed (black), $u^\prime$; 
    \solid (blue), $v^\prime$;
    \chndot (red), $w^\prime$.
  }
  \label{fig:urms_33} 
%
\vspace{3mm}
  \centering
  \includegraphics[width=0.8\linewidth,clip]{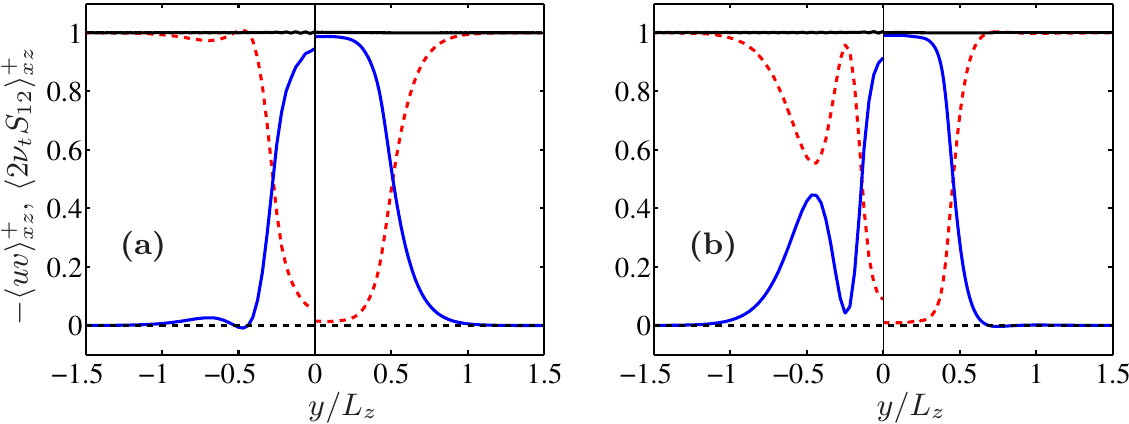}
  \caption{(a,b) As in figure \ref{fig:urms_33}, for the momentum balance.
    \solid (blue), $\langle -u v\rangle_{xz}/u_\tau^2$;
    \dashed (red), $\langle  2\nu_t \sbar_{xy} \rangle_{xz}/u_\tau^2$; 
    \solid (black), total stress.
  }
  \label{fig:momentum_uv_33} 
\end{figure}

\section{Vertical localised upper-branch equilibria, and localised turbulence}\label{sec:tall-EQ}

\subsection{The effect of box aspect ratios and characterisation of equilibria}\label{sec:box-dependency}

Figure~\ref{fig:LES_bif_box_dependency}(a) shows the dependence on $A_{yz}$ of the Reynolds
stress at the central plane for $R_S=38.9$ and $A_{xz}=3$, close to the initial bifurcation.
Solutions only exist for $A_{yz} > 1.1$. The Reynolds stress decreases as $A_{yz}$
increases, and may approach a non-zero constant in the limit $A_{yz} \rightarrow \infty$.
The same is true for the velocity fluctuations (not shown). Note that, since the grid
becomes relatively coarser as $A_{yz}$ increases, a finer $y$ grid is used for taller boxes
(dashed lines), and that the good agreement of the results whenever the two grids overlap
confirms numerical convergence.

The dependence on the streamwise aspect ratio $A_{xz}$ is shown in
figure~\ref{fig:LES_bif_box_dependency}(b). Solutions exist for $1.58 \lesssim A_{xz}
\lesssim 3.29$, which covers the range of box aspect ratios ($A_{xz} \approx 3$) identified
by \citet{SekimotoDongJimenez2016} to be good models for wall-bounded shear turbulence. It
is interesting that the minimum aspect ratio for steady Nagata equilibrium solutions in
plane Couette flow is $A_{xz}\approx 1.62$ at low Reynolds numbers
\citep{JimenezKawaharaSimensNagataShiba2005}, and that \citet{Deguchi2015} showed that they
only exist $A_{xz} \gtrsim 1.5$, even at high Reynolds numbers. On the other hand
\cite{KawaharaKida2001} found unstable periodic orbits (UPOs) for somewhat shorter boxes,
$A_{xz} =1.46$.

Figure~\ref{fig:LES_bif_Ax3_invCz} shows the continuation along $R_S$ for $A_{xz}=3$ and
several $A_{yz}$. The box-averaged statistics of LES (symmetric) turbulence are also shown
for comparison. For $R_S \lesssim 60$, those turbulence simulations occasionally become
vertically localised around $y=0$, but they spread again to fill the whole domain (see
\S\ref{sec:local-turb}). The box-averaged turbulence statistics in
figure~\ref{fig:LES_bif_Ax3_invCz} include such locally quiescent regions, leading to weaker
$v^\prime$ and $w^\prime$ fluctuations than would otherwise be obtained at the central
plane. When $R_S \lesssim 50$, LES turbulence often decays to laminar after these
localisation events.

The velocity fluctuations for the upper-branch solutions are quite large in flat boxes
 and low Reynolds number,  $A_{yz}=1.5$ and $R_S \approx 45$, but that behaviour disappears
for taller boxes and saturates beyond $A_{yz}\approx 3$. These large fluctuations are
thus probably an effect of the shear-periodic boundary condition. This is also the
range in which the shear-periodicity results in the strongest temporal oscillations of the
kinetic energy, of the order of 1\%, but we tested that the fluctuations in
figure~\ref{fig:LES_bif_Ax3_invCz}(a, c) are not due to the temporal variability. They are also
present in the spacial average of instantaneous snapshots. The temporal oscillation of
solutions with $A_{yz} > 2$ is of the order of $10^{-5}$.
 
The relatively poor scaling of the velocity fluctuations of the LES equilibria with $SL_z$
can be traced to the poor scaling of $|\sigma|/S$. Figure
\ref{fig:LES_bif_Ax3_invCz}(d) shows that the dimensionless strain rate of the upper-branch
equilibria is roughly unity for $R_S = 50$--$100$, while that of turbulence is in the range
of $2$--$5$. It was shown by \citet{SekimotoDongJimenez2016} that, although $SL_z$ is the
natural velocity scale for `good' DNS boxes, the fluctuations in non-optimal boxes
scale better with the friction velocity obtained from their total measured stress. In
essence, the Reynolds stress and the velocity fluctuations scale with each other.

The same is true for LES and for equilibrium solutions. Figures
\ref{fig:LES_bif_Ax3_invCz_wall_unit}(a,c,d) show that the fluctuations collapse to a common
curve in these `wall' units. The vorticity components also scale well with $u_\tau/l_S$ (not
shown). In this normalisation, the velocity fluctuations of symmetric LES turbulence at
$R_S=100$ are $u^{\prime +} \approx 2$, $v^{\prime +} \approx 1.2$ and $w^{\prime +} \approx 1.4$, 
which agrees well with those at the top of the logarithmic layer of turbulent channels. 
When $R_S< 60$, the velocity fluctuations of the LES equilibria are not very different from those of
turbulence, but we have seen that the lower-branch solutions tend to get concentrated around
the critical layer as $R_S$ increases. Their statistics then become very different from
turbulence.

Figure~\ref{fig:LES_bif_Ax3_invCz_wall_unit}(b) is an enlargement of figure
\ref{fig:LES_bif_Ax3_invCz_wall_unit}(a), showing the dependence on $A_{xz}$ of the minimum
bifurcation Reynolds number of equilibria with $A_{yz}=3$. It turns out that the bifurcation
is more dependent on $A_{xz}$ than on $A_{yz}$. At a fixed $A_{xz}=3$, the minimum
$R_S$ is $(38.11, 38.00, 37.94, 37.91, 37.89, 37.90)$ for $A_{yz} = (1.33, 1.5, 1.64, 1.8,
2.0, 3.0)$, with a maximum scatter of 0.6\%. On the other hand, when $A_{yz}$ is fixed at 3,  $R_S$ 
changes by 8\%, from 35.05 to 37.90, as $A_{xz}$ changes from 2 to 3.

\begin{figure}
  \centering
  \includegraphics[width=0.8\linewidth,clip]{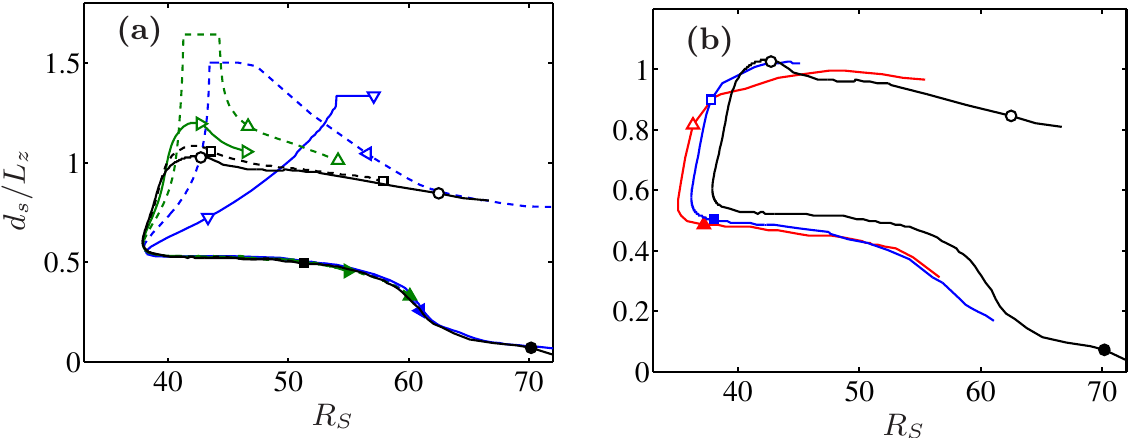}
\caption{Height of the velocity streak, defined by the distance $d_s$ between the peaks of
$u^\prime$ in the core part of the solutions in figure \ref{fig:urms_33}. Only peaks in the
inner core are considered, and the secondary peaks in the outer one-component layer are
ignored.
(a) $A_{xz}=3$ and $A_{yz}=1.33,~1.5,~1.64,~1.8,~2,~3$. Lines and symbols as in
figure~\ref{fig:LES_bif_Ax3_invCz}.
(b) $A_{xz}=2,2.5,3$ and $A_{yz}=3$. Lines and symbols as in
figure~\ref{fig:LES_bif_Ax3_invCz_wall_unit}(b).
  }
  \label{fig:height_yd} 
\end{figure}

\begin{figure}
  \centering
  \includegraphics[width=0.85\linewidth,clip]{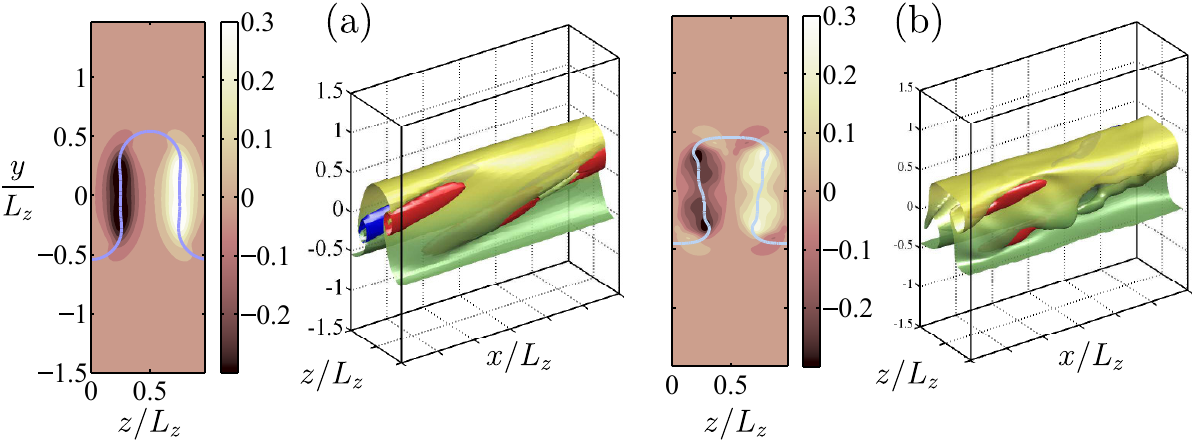}
  \caption{
    Vertically localised upper-branch solutions: 
    As in figure~\ref{fig:UPO_LES_flat_3d}, for $A_{xz}=3,A_{yz}=3$ and: (a) $R_S = 42.7$ and (b) $R_S =62.5$. 
  }
  \label{fig:ox_u} 
  \vspace{2mm}
  \centering
  \includegraphics[width=0.7\linewidth,clip]{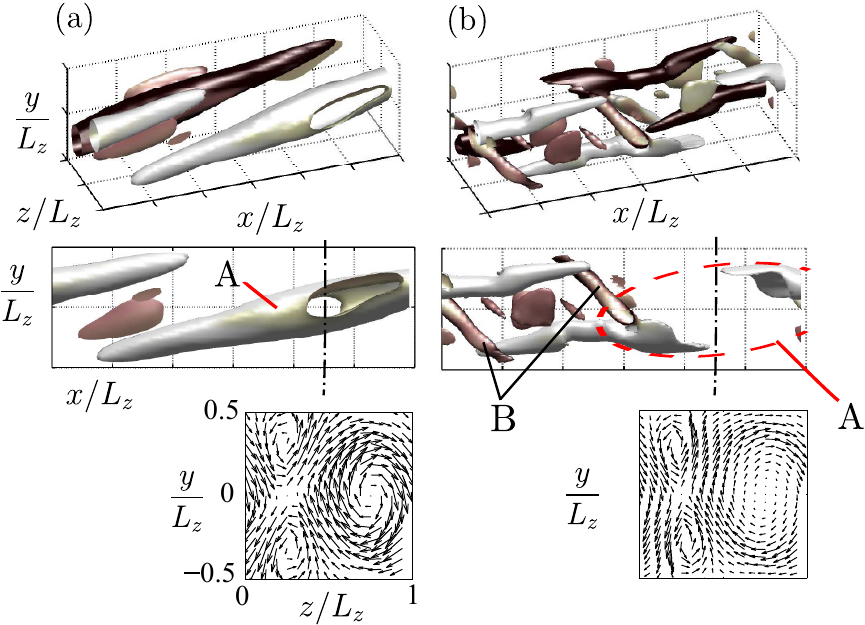}
\caption{%
Isosurfaces of the second invariant of the velocity gradient ($Q=0.2Q_\mathrm{max}$)
coloured by $\omega_x $, with the same colour code as in the streamwise-averaged 
$\omega_x$ maps in the left parts of figures~\ref{fig:ox_u}(a,b):
(dark-grey) $\omega_x < 0$, (light-grey) $\omega_x > 0$. 
From top to bottom, each figure shows: a three-dimensional view, a side view of the
region $0.5 \leq z/L_z \leq 1$, and a cross section of cross-flow velocity vectors at $x/L_z =
2.25$, marked as a chain-dotted vertical line in the side views. The primary streamwise rollers 
visible in these cross-sections are not always compact enough to appear in the Q isosurfces and are  marked as (A) in the lateral views.
Flow as in figures \ref{fig:ox_u}(a,b). Only the active part, $|y|/L_z <0.5$, is shown in all cases.
  }
\label{fig:qcrit} 
  \vspace{2mm}
  \centering
  \includegraphics[width=0.35\linewidth,clip]{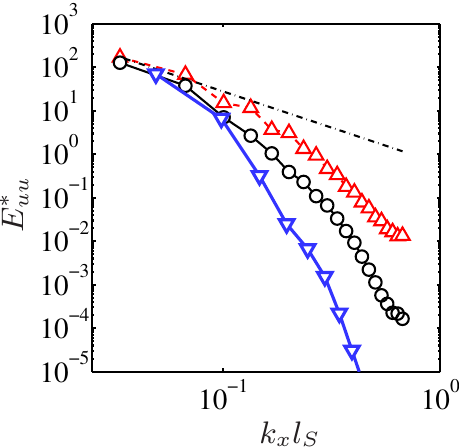}
  \caption{Streamwise velocity spectrum $E_{uu}^\ast = E_{uu}(k_x) \varepsilon^{-2/3} l_S^{-5/3}$ as function of $k_x l_S$: 
    $\triangledown$ (blue), upper-branch solution at $R_S=42.7$; 
    $\circ$ (black), upper-branch solution at $R_S=62.5$; 
    $\triangle$ (red), turbulence LES  at $R_S=62.5$. 
    The dash-doted line represents the inertial theory $E_{uu}^\ast(k_x)=0.6  (l_S k_x)^{-5/3}$.
    The spectra and dissipation rates of the upper-branch solutions are averaged over $|y|/L_z < 0.5$.
  } 
  \label{fig:spec} 
\end{figure}


\subsection{The structure of the upper-branch equilibria}\la{sec:upper-branch}

We focus next on the flow structure of the vertically localised upper-branch equilibria in a
box with $A_{xz}=3$, $A_{yz}=3$. The right-hand side of the two panels in
figure~\ref{fig:urms_33} shows that the velocity fluctuations of these solutions decay
exponentially away from $y=0$, which is a common feature of localised solutions in other
shear flows~\citep{SchneiderGibsonBurke2010,GibsonBrand2014}. The tail of $u^\prime$ is much
stronger than those of $v^\prime$ and $w^\prime$. This can be shown to be due to a
streamwise-constant `streak' of the streamwise velocity that can be approximated as a
roughly sinusoidal spanwise perturbation of the mean profile, $u_{s} \sim \sin (2\pi
z/L_z)$. This weak perturbation is the far tail of the stronger sinuous streak of $u$
concentrated near $y=0$ and, because it is almost independent of $x$, is essentially
independent of $v$, $w$ and of the shear. The left-hand part of both panels in
figure~\ref{fig:urms_33} represent lower-branch solutions, which share many of the
characteristics of the upper branch. As the Reynolds number increases, all structures
become more complex, as seen in figure~\ref{fig:urms_33}(b) and later in figures
\ref{fig:ox_u} and \ref{fig:qcrit}, and the core of the structures develop substructures
that could perhaps be interpreted as a first indication of a turbulent cascade. The momentum
balance of these flows is shown in figure~\ref{fig:momentum_uv_33}. The Reynolds stress is
dominant in the core, $|y|/L_z < 0.5$, but only the mean shear $\dr U / \dr y $ contributes
to the eddy viscosity in the outer part.

Figure~\ref{fig:height_yd} presents the height of the velocity streak for the different
equilibria, defined as the twice the distance to $y=0$ of the first maximum of the $u^\prime$ 
profile of the solutions in figure \ref{fig:urms_33}. The height of lower-branch streak
stays roughly constant below $ R_S \approx 50$, and approaches zero quickly above that
limit, as the solutions collapse to the critical layer. In contrast, the height of
upper-branch solutions increases drastically near the bifurcation point and reaches a
maximum near $R_S \approx 45$. This is also the range in which the velocity fluctuations
become very strong in figure~\ref{fig:LES_bif_Ax3_invCz}, and only appears in flat boxes
with $A_{yz}=1.33-1.64$. We argued in the discussion of figure~\ref{fig:LES_bif_Ax3_invCz}
that upper-branch solutions tend to be limited by the height of these flat boxes, and
figure~\ref{fig:height_yd}(a) confirms this interpretation by showing that the maximum
height of the streak reaches the box height. On the other hand, this interaction does not
take place when $A_{yz} \gtrsim 2$, confirming the independence of those solutions from the
box dimensions. 

The flow structures of the upper-branch solutions at $R_S=42.7$ and $62.5$ are visualised in
figures~\ref{fig:ox_u} and \ref{fig:qcrit}. The streak is represented by the $\ubar=0$
isosurface that spans $|y|/L_z< 0.5$ in figure~\ref{fig:ox_u} (see also
figure~\ref{fig:UPO_LES_flat_3d}a,b). The corresponding vortical structures are located in
the flanks of the streak, as seen in the streamwise-averaged cross-stream maps of $\omega_x$
in figure~\ref{fig:ox_u}. The vortical structures are shown by themselves in
figure~\ref{fig:qcrit}, and are surprisingly complex for an equilibrium solution. This is
especially true for the higher Reynolds number case in figures~\ref{fig:ox_u}(b) and
\ref{fig:qcrit}(b), which appear to include a double structure that is nevertheless in
equilibrium. In fact, there is a first indication of two separate scales in these flow
fields. The smaller tube-like vortical structures are isosurfaces of the second invariant of
the velocity gradient tensor (Q criterion), and are coloured by the streamwise vorticity.
They do not always coincide with the larger-scale structures of the cross-stream velocity,
which are visible in the cross sections in the bottom part of figure \ref{fig:qcrit}. These
`rollers' are too diffuse to appear in the Q-map, which may actually appear empty in the
region in which the roller is dominant (see the region labelled as (A) in the side view in
figure \ref{fig:qcrit}b, and note that, because of the symmetry of the flow, a
mirror-symmetric arrangement is present in the upstream half of the box). On the
other hand, it is clear from the cross sections that the roller dominates the velocity
field.

The low-Reynolds number solution in figure~\ref{fig:qcrit}(a) shows that the Q-vortices are
parts of the larger streamwise rollers that have been sheared and stretched by the mean
flow. The cross sections in the bottom part of figure~\ref{fig:qcrit} shows that these
vortices are long enough to spill into neighbouring periodic boxes, so that they appear as
double in the cross section. The inclination angle of these streamwise rollers and vortices
is roughly 15$^\circ$ at all Reynolds numbers. In the higher-Reynolds number case in
figure~\ref{fig:qcrit}(b) the streamwise vortices are strong enough to create new
vortices, roughly perpendicular to them, which are labelled as (B) in the side view in
figure~\ref{fig:qcrit}(b). They rotate in the opposite sense to the primary streamwise
rollers and are aligned in the direction of the strain produced by them. A similar
generation mechanism of secondary smaller vortices has been investigated in the homogeneous
isotropic turbulence~\citep{Goto2012}.

The velocity spectra in figure~\ref{fig:spec} show that the upper-branch solutions acquires
more small-scale structures as $R_S$ increases, approaching the spectrum of turbulence LES
at similar Reynolds numbers. Even though the turbulence state has more small scale, the
large-scale end of its spectrum is quantitatively similar to the upper-branch solutions.

%
\begin{figure} 
  \centering
  \begin{minipage}{0.9\linewidth}
    \includegraphics[width=1.\linewidth,clip]{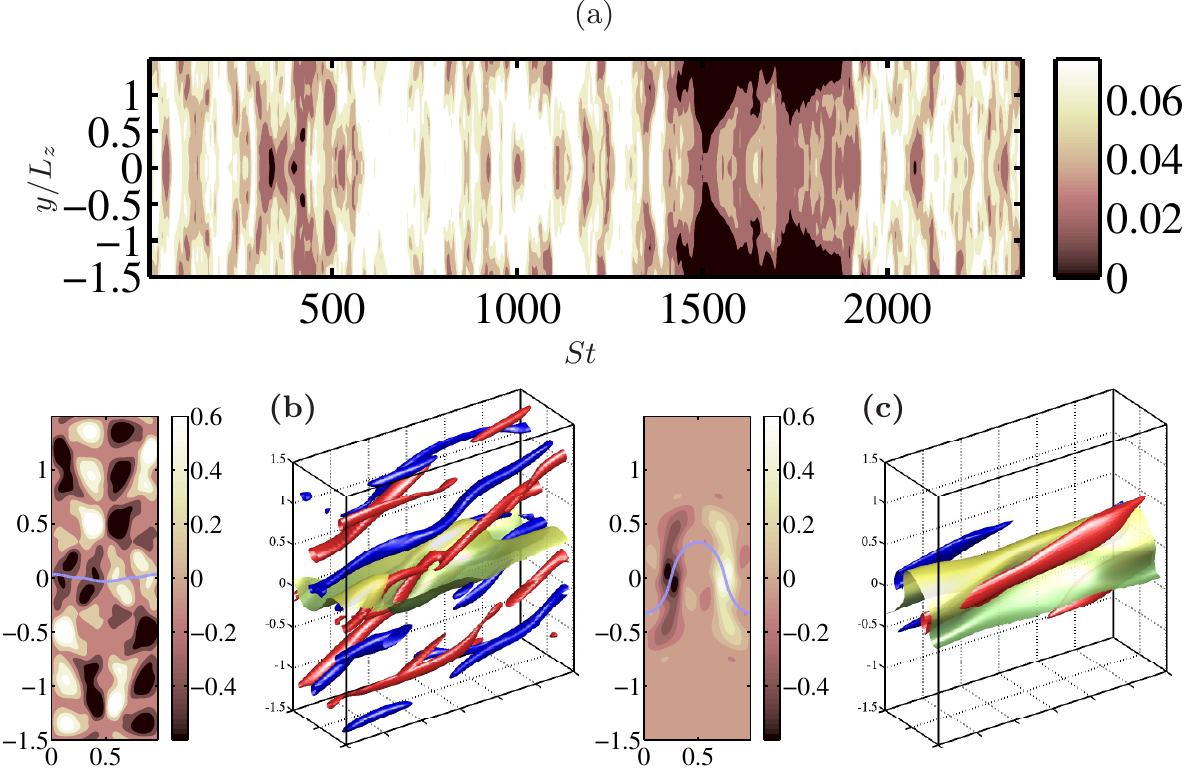} 
  \end{minipage} 
\caption{%
(a) Vertical velocity fluctuation intensity profiles, $\langle v^{2}
\rangle_{xz}^{1/2}/(SL_z)$, as a function of $y/L_z$ and $St$. The contours are
[0.02:0.02:0.08], and the mean value of $v^\prime$ is approximately 0.07 $SL_z$ (see
figure~\ref{fig:LES_bif_Ax3_invCz}). $A_{xz}=3$, $A_{yz}=3$, $R_S = 50$.
(b) Homogeneous turbulence at $St=1318$, 
(c) Localised state at $St=1525$. 
The left panels in (b,c) show the streamwise-averaged $\omega_x$, and the right ones show
isosurfaces of $\omega_x$ and $\ubar$, as in figure~\ref{fig:UPO_LES_flat_3d}.
  }
  \label{fig:LES_localised_turb} 
\end{figure}

\begin{figure}
  \centering
  \includegraphics[width=0.9\linewidth,clip]{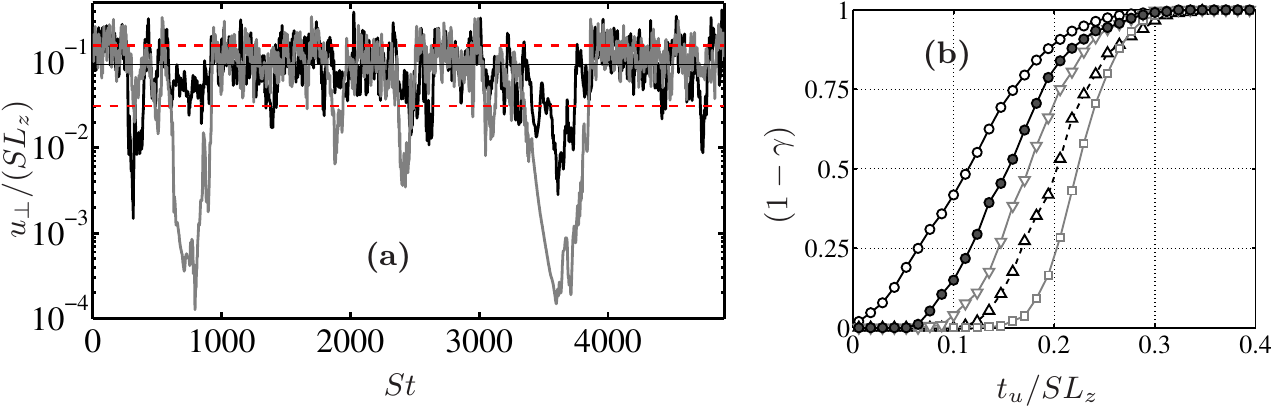}
  \caption{%
    (a) The local cross-flow velocity fluctuation intensity $u_\perp$ defined in (\ref{eq:ucross})
    at: (black) $y=L_y/2$ ($u_t$), (grey) $y/L_z=0$ ($u_c$). The black dashed line is the mean value, 
    $\bra u_\perp \ket$, and the red dashed lines are $ \langle u_\perp \rangle \pm \sigma$, where
    $\sigma$ is the standard deviation. They are defined as averages over the centre and top of
    the box, as explained in the text.
    (b) The intermittency factor $(1 - \gamma)$, where $\gamma$ is defined in (\ref{eq:intermittency}): \solidcirc, $R_S = 50.0$; \solidcircf, $R_S = 52.6$; \solidtridown (grey), $R_S=55.5$; \dashedtri, $R_S = 62.5$; \solidsquare (grey), $R_S = 101.6$. 
  }
  \label{fig:LES_localised_freq_pdf} 
\end{figure}

\subsection{Intermittent visiting of turbulence to vertically-localised states}\la{sec:local-turb}

We mentioned in \S\ref{sec:box-dependency} that, when $R_S$ is relatively low, LES turbulence
collapses intermittently to a localised state around $y=0$, and that these states persist for a
long time when $R_S \gtrsim 50$. These are the statistics plotted as diamonds in
figure~\ref{fig:LES_bif_Ax3_invCz}. Figure~\ref{fig:LES_localised_turb}(a) shows the temporal 
evolution of the profile of local $v^\prime$. Light colours represent active turbulent
regions and dark ones are overdamped `laminar' areas. Localised turbulence occurs when
laminarisation does not extend over the whole height of the box, such as in $St=1500-2000$.

Note that localisation only takes place in flows with sinuous symmetry, so that the profiles
are symmetric with respect to $y=0$, but that the symmetry allows localisation both at the
centre plane, $y=0$, or at the top or bottom of the box, $y=\pm L_y/2$. The strongest event
in figure~\ref{fig:LES_localised_turb}(a) is localised at $y=0$ $(St=1500-2000)$, but there
are several minor ones (e.g., at $St\approx 400$) localised at the top of the box.

The structure of the flow at a fully turbulent moment is shown in
figure~\ref{fig:LES_localised_turb}(b), and shows that vorticity is spread over the whole
box. Figure~\ref{fig:LES_localised_turb}(c) represents the flow during a localisation event.
Its similarity to the localised lower-branch equilibrium solutions is striking. Since we saw in
\S\ref{sec:overview} that these solutions are edge states whose unstable manifold leads to
bursting, it is tempting to interpret localisation as the occasional approach of
fully-developed turbulence to the localised solution along its stable manifold, and its later
spreading to a fully turbulent state along the unstable manifold. A similar behaviour was
reported in a plane channel by~\cite{ItanoToh2001}, and \citet{kawahara05} used the
occasional approaches to the edge state to laminarise turbulent Couette flow.

We next quantify the frequency of these localisation events.
Figure~\ref{fig:LES_localised_freq_pdf}(a) shows the evolution of the transverse velocity
fluctuation intensity, defined as
\beq
u_\perp(y) \equiv \left(\langle v^2 \rangle_{xz} +\langle w^2 \rangle_{xz}\right )^{1/2}.
\la{eq:ucross}
\eeq
The darker line is the intensity at $y=0$, and the lighter one is the intensity at $y = \pm
L_y/2$. The thin straight solid and the dashed lines are the mean, 
$\bra u_\perp \ket$, and deviation, $\sigma$, which are the averages of these two positions.
By defining $u_c=u_\perp(0)$, and $u_t=u_\perp(L_y/2)$, fully turbulent profiles should have large $u_c\approx u_t \approx \bra u_\perp \ket$, while localised states are characterised by having one of these
intensities much weaker than the fully turbulent one. 
The quantity
\begin{equation}\label{eq:intermittency}
\gamma \equiv \int_{t_u}^{\infty} \int_{t_u}^{\infty} P(u_c,u_t) \dd u_c \dd u_t,  
\end{equation}
where $P(u_c,u_t)$ is the joint probability density function of both intensities, 
and $t_u$ is a threshold, measures the probability that the points $y=0$ or $L_y/2$ are turbulence according to $u_c \geq t_u$ or $u_t \geq t_u$.
Since the localised state has the overdamped laminar which intermittently appears as shown in figure~\ref{fig:LES_localised_freq_pdf}(a), $(1 - \gamma)$ indicates the intermittency factor of the localised turbulence.
Figure~\ref{fig:LES_localised_freq_pdf}(b) shows that the frequency of localisation decreases as  $R_S $ increases. 

%
\section{Conclusions}\la{sec:conc}

We have performed large-eddy simulations of statistically stationary homogeneous shear
turbulence (SS-HST) in a subspace with sinuous symmetry, and found equilibrium solutions
with the same symmetry. We use a Smagorinsky-type model with no molecular viscosity, which
is not required because of the absence of walls, so that the eddy-viscosity $\nu_t\equiv
l_S^2 |S| $ acts as the only energy sink. It is parametrised by a mixing length $l_S$ that
plays the role of the Kolmogorov scale in LES, independent of the numerical grid. For the
grids used in this study, the flow is independent of the numerical resolution. The integral
scale in the LES of the SS-HST is comparable to the spanwise box dimension, $L_0 \approx 0.4
L_z$, as in the DNS of the same flow in \cite{SekimotoDongJimenez2016} and as in
wall-bounded turbulence in spanwise-limited boxes \citep{FloresJimenez2010}. The effective
Kolmogorov scale is $\eta_t \approx 0.9 l_S$, and the velocity fluctuations scale well with
$SL_z$, as in DNS. Even if we introduce the molecular viscosity in our LES, considering the effective Kolmogorov length, $\widetilde{\eta_t} \approx \widetilde{l_S} \equiv l_S\left( 1 + \nu / \nu_t \right)^{1/2} $ (see \cite{pope00book}), the results would not change as long as $\nu$ is enough small with respect to the mean $\nu_t$.

The length-scale ratio $R_S = L_z/l_S$ is used as a continuation parameter for LES
equilibria, playing the role of a Reynolds number, and it is found that vertically localised
equilibrium solutions appear by a saddle-node bifurcation at $R_S=37.9$ for $A_{xz}=3$. The dependence of the equilibrium solutions on the
box aspect ratios has been investigated, and it is revealed that both lower- and
upper-branch solutions tend to localise vertically around the central plane of box. The initial bifurcation point is roughly
independent of $A_{yz}$, and much more dependent on $A_{xz}$. These solutions exist in $1.58
\lesssim A_{xz} \lesssim 3.29$ and $A_{yz}>1.1$ at $R_S=39.0$. This range of aspect ratios
spans those found by \cite{SekimotoDongJimenez2016} to be good models for unconstrained shear
flows in general. The minimum limit of $A_{xz} \gtrsim 1.6$ is also similar to those
found for the existence of equilibria in plane Couette flow by \cite{Nagata1990} and
\cite{Deguchi2015}. These and other authors have mentioned that such equilibria can be
embedded in general unbounded shear flows, but the localisation of the present solution in
an approximately linear shear is almost surely due to the interaction with the local eddy
viscosity profile. As $R_S$ increases, the lower-branch solutions take the form of a
critical layer, such as those found in previous works on wall-bounded
flows~\citep{WangGibsonWaleffe2007,Viswanath2009,DeguchiHall2014JFM_PCF}, and described by vortex-wave
interaction (VWI) theory~\citep{HallSmith1991,HallSherwin2010}.

The length scales of the LES equilibria are $l_S$ for the small scale and $L_z$ for the
large one, as in LES turbulent. The comparison of the contribution of the eddy viscosity and
of the Reynolds stress to the momentum balance reveals that the former is weak within
the localised equilibrium solutions, and predominate outside it. 
The velocity fluctuations of the
present equilibria are substantially smaller than those of self-sustaining turbulence, especially in
the VWI limit, and do not scale well with $SL_z$. However, they
scale well with their own $u_\tau$, with similar values to those of wall-bounded flows
expressed in wall-units.

At low Reynolds numbers, lower-branch solutions act as edge states. Although their
eigenvalue structure is more complicated than a simple saddle, one of the unstable
directions of the saddle leads to an exponential burst and to chaotic turbulence, while the
other laminarises. In turbulent LESes, the flow occasionally collapses to a localised state
which resembles the equilibrium solutions. Depending on the Reynolds number, the outcome of
these events is more often reinjection to turbulence, or laminarisation.
 
Upper-branch solutions have tall velocity streaks associated with small-scale vortices,
whose complication increases with increasing $R_S$. It is interesting that, even at the
relatively low $R_S\approx 62$, small secondary vortices begin to appear in these solutions,
aligned perpendicularly to the primary streamwise rollers in a manner strongly reminiscent
of the multiscale process frequently invoked as models for the turbulent cascade.
Further continuations to higher $R_S$ are hardly successful probably because of 
their increasing complexity and instability, 
which is the similar limitation to the one we encounter when searching dynamically important invariant solutions in the Navier-Stokes computations at high Reynolds numbers.

In all, the localised LES equilibria discovered here represent a promising model for generic isolated 
turbulent structures in shear flows. Most intriguingly, the higher Reynolds numbers contain what
appear to be the first stages of a multiscale cascade. 

\section*{Acknowledgements}
This research has been funded by the European Research Council grants ERC-2010.AdG-20100224
and ERC-2014.AdG-669505. We are grateful to G. Kawahara for early discussions.

\vspace{1cm}

\appendix


\section{Linear stability analysis of equilibria in LES}~\label{appdx:floquet}

\begin{figure}
  \centering
  \includegraphics[width=0.9\linewidth,clip]{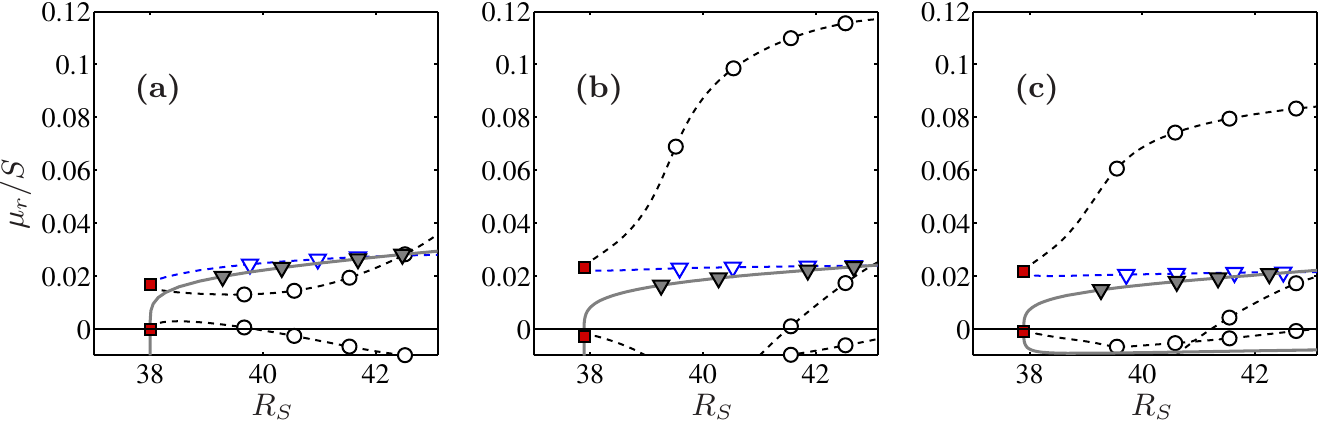}
  \caption{The real-part of nondimensional eigenvalues obtained by the linear stability analysis of equilibria for $A_{xz}=3$ and (a) $A_{yz}=1.5$, (b) $A_{yz}=2$, (c) $A_{yz}=3$. 
    $\triangledown$, lower branch; $\circ$, upper branch. The filled (open) symbols represent real (complex conjugate) modes.
    Positive values represent unstable modes, and the square is the bifurcation point.
  }
  \label{fig:arnoldi_Ax3y3} 
\end{figure}

We discuss in this section the linear stability of the vertically localised symmetric
equilibria described in the body of the paper. Even if we saw in
figure~\ref{fig:LES_bif_Ax3_invCz_wall_unit}(a) that the continuation diagram of these
solutions is roughly independent of the vertical box aspect ratio, it turns out that this
aspect ratio affects the stability of the equilibria, especially that of the upper-branch
solutions. Since we showed in figure~\ref{fig:height_yd} that these solutions are tall
enough to span the full height of the box, especially in the case of the flatter boxes with
$A_{yz} \le 1.5$, the artificial interactions with the shear-periodic copies in $y$ is
inevitable. Figure~\ref{fig:arnoldi_Ax3y3} shows the distribution of a few of the least
stable eigenvalues obtained from the linear stability analysis of equilibria with $A_{xz}=3$
and $A_{yz}=1.5, 2$ and $3$. The vertically localised equilibrium solution appears through a
saddle-node bifurcation as $R_S$ increases. The upper branch always has at least one
pair of unstable complex conjugate modes, and becomes more stable as $A_{yz}$ decreases.

In taller boxes, with $A_{yz}=2-3$, the distribution of unstable eigenvalues of the
upper-branch solutions is not strongly affected by the box aspect ratio, in rough agreement
with the criterion, $A_{xz} \lesssim 2 A_{yz} $, found in previous
DNS studies of SS-HST to be required  for box independence \citep{SekimotoDongJimenez2016}.

\bibliographystyle{jfm}
\bibliography{database}

\end{document}

%% file: set_newcommand.tex
\newcommand{\solid}{---$\!$---}
\newcommand{\solidcirc}{---$\circ$---}

\newcommand{\solidsquare}{---$\square$---}
\newcommand{\solidcircf}{---$\bullet$---}

\newcommand{\solidtridown}{---$\triangledown$---}

\newcommand{\solidtriright}{---$\triangleright$---}
\newcommand{\dashed}{\hbox{{--}\,{--}\,{--}\,{--}}}
\newcommand{\chndot}{---\,$\cdot$\,---}

\newcommand{\dashedcircf}{\hbox{{--}\,{--}\,$\bullet${--}\,{--}}}

\newcommand{\dashedtri}{\hbox{{--}\,{--}\,$\vartriangle${--}\,{--}}}
\newcommand{\dashedtridown}{\hbox{{--}\,{--}\,$\triangledown${--}\,{--}}}

  \newcommand{\itbold}[1]{\textbf{\textit{#1}}}

  \newcommand{\xvec}{\itbold{x}}

  \newcommand{\fvec}{\itbold{f}}

\newcommand{\ubar}{\overline{u}} 
\newcommand{\pbar}{\overline{p}} 
\def\dd{{\, \rm{d}}}
\def\dr{{\rm{d}}}

\def\bra{\langle}
\def\ket{\rangle}

\def\beq{\begin{equation}}
\def\eeq{\end{equation}}
\def\la{\label}

\def\ii{{\rm i}}

\def\dis{\varepsilon}

\def\sbar{\overline{\sigma}}

\def\ombar{\overline{\omega}}

%% file: les_hst_sekimoto_arXiv.bbl
\begin{thebibliography}{62}
\expandafter\ifx\csname natexlab\endcsname\relax\def\natexlab#1{#1}\fi

\bibitem[Avila {\em et~al.\/}(2013)Avila, Mellibovsky, Roland \&
  Hof]{AvilaMellibovskyRolandHof2013}
{\sc Avila, M., Mellibovsky, F., Roland, N. \& Hof, B.} 2013
  Streamwise-localized solutions at the onset of turbulence in pipe flow. {\em
  Phys. Rev. Lett.\/} {\bf 110}, 224502.

\bibitem[Blackburn {\em et~al.\/}(2013)Blackburn, Hall \&
  Sherwin]{BlackburnHallSherwin2013}
{\sc Blackburn, H.~M., Hall, P. \& Sherwin, S.} 2013 Lower branch equilibria in
  couette flow: the emergence of canonical states for arbitrary shear flows.
  {\em J. Fluid Mech.\/} {\bf 726}, R2.

\bibitem[Cambon \& Scott(1999)]{CambonScott1999}
{\sc Cambon, Claude \& Scott, Julian~F.} 1999 Linear and nonlinear models of
  anisotropic turbulence. {\em Ann. Rev. Fluid Mech.\/} {\bf 31}, 1--53.

\bibitem[Champagne {\em et~al.\/}(1970)Champagne, Harris \&
  Corrsin]{champ:h:c:70}
{\sc Champagne, F.H., Harris, V.G. \& Corrsin, S.} 1970 Experiments on nearly
  homogeneous turbulent shear flow. {\em J. Fluid Mech.\/} {\bf 41}, 81--139.

\bibitem[Chandler \& Kerswell(2013)]{ChandlerKerswell2013}
{\sc Chandler, Gary~J. \& Kerswell, Rich~R.} 2013 Invariant recurrent solutions
  embedded in a turbulent two-dimensional kolmogorov flow. {\em J. Fluid
  Mech.\/} {\bf 722}.

\bibitem[Cvitanovi{\'c}(2013)]{Cvitanovic2013}
{\sc Cvitanovi{\'c}, P.} 2013 Recurrent flows: the clockwork behind turbulence.
  {\em J. Fluid Mech.\/} {\bf 726}, 1--4.

\bibitem[Deguchi(2015)]{Deguchi2015}
{\sc Deguchi, K.} 2015 Self-sustained states at kolmogorov microscale. {\em J.
  Fluid Mech.\/} {\bf 781}, R6.

\bibitem[Deguchi \& Hall(2014{\natexlab{{\em a\/}}})]{DeguchiHall2014_PSTA}
{\sc Deguchi, K. \& Hall, P.} 2014{\natexlab{{\em a\/}}} Canonical exact
  coherent structures embedded in high {R}eynolds number flows. {\em Phil.
  Trans. R. Soc. A\/} {\bf 372}, 20130352.

\bibitem[Deguchi \& Hall(2014{\natexlab{{\em b\/}}})]{DeguchiHall2014JFM_PCF}
{\sc Deguchi, K. \& Hall, P.} 2014{\natexlab{{\em b\/}}} The
  high-{R}eynolds-number asymptotic development of nonlinear equilibrium states
  in plane couette flow. {\em J. Fluid Mech.\/} {\bf 750}, 99--112.

\bibitem[Deguchi \& Hall(2016)]{DeguchiHall2016}
{\sc Deguchi, K. \& Hall, P.} 2016 On the instability of vortex-wave
  interaction states. {\em J. Fluid Mech.\/} {\bf 802}, 634--666.

\bibitem[Deguchi {\em et~al.\/}(2013)Deguchi, Hall \&
  Walton]{DeguchiHallWalton2013}
{\sc Deguchi, K., Hall, P. \& Walton, A.} 2013 The emergence of localized
  vortex-wave interaction states in plane {Couette} flow. {\em J. Fluid
  Mech.\/} {\bf 721}, 58--85.

\bibitem[Faisst \& Eckhardt(2003)]{FaisstEckhardt2003}
{\sc Faisst, H. \& Eckhardt, B.} 2003 Traveling waves in pipe flow. {\em Phys.
  Rev. Lett.\/} {\bf 91}, 224502.

\bibitem[Flores \& Jim\'enez(2010)]{FloresJimenez2010}
{\sc Flores, O. \& Jim\'enez, J.} 2010 Hierarchy of minimal flow units in the
  logarithmic layer. {\em Phys. Fluids\/} {\bf 22}, 071704.

\bibitem[Gerz {\em et~al.\/}(1989)Gerz, Schumann \&
  Elghobashi]{GerzSchumannElghobashi1988}
{\sc Gerz, T., Schumann, U. \& Elghobashi, S.~E.} 1989 Direct numerical
  simulation of stratified homogeneous turbulent shear flows. {\em J. Fluid
  Mech.\/} {\bf 200}, 563--594.

\bibitem[Gibson \& Brand(2014)]{GibsonBrand2014}
{\sc Gibson, J.~F. \& Brand, E.} 2014 Spanwise-localized solutions of planer
  shear flows. {\em J. Fluid Mech.\/} {\bf 745}, 25--61.

\bibitem[Goto(2008)]{Goto2008}
{\sc Goto, S.} 2008 A physical mechanism of the energy cascade in homogeneous
  isotropic turbulence. {\em J. Fluid Mech.\/} {\bf 605}, 355--366.

\bibitem[Goto(2012)]{Goto2012}
{\sc Goto, S.} 2012 Coherent structures and energy cascade in homogeneous
  turbulence. {\em Prog. Theor. Phys. Suppl.\/} {\bf 195}, 139--156.

\bibitem[Gualtieri {\em et~al.\/}(2007)Gualtieri, Casciola, Benzi \&
  Piva]{Gualtieri2007}
{\sc Gualtieri, P., Casciola, C.~M., Benzi, R. \& Piva, R.} 2007 Preservation
  of statistical properties in large-eddy simulation of shear turbulence. {\em
  J. Fluid Mech.\/} {\bf 592}, 471--494.

\bibitem[Hall \& Sherwin(2010)]{HallSherwin2010}
{\sc Hall, P. \& Sherwin, S.} 2010 Streamwise vortices in shear flows:
  harbingers of transition and the skeleton of coherent structures. {\em J.
  Fluid Mech.\/} {\bf 661}, 178--205.

\bibitem[Hall \& Smith(1991)]{HallSmith1991}
{\sc Hall, P. \& Smith, F.T.} 1991 On strongly nonlinear vortex/wave
  interactions in boundary-layer transition. {\em J. Fluid Mech.\/} {\bf 227},
  641--666.

\bibitem[Hughes {\em et~al.\/}(2001)Hughes, Oberai \&
  Mazzei]{HughesOberaiMazzei2001}
{\sc Hughes, T. J.~R., Oberai, A.~A. \& Mazzei, L.} 2001 Large eddy simulation
  of turbulent channel flows by the variational multiscale method. {\em Phys.
  Fluids\/} {\bf 13}, 6.

\bibitem[Hwang \& Cossu(2010)]{HwangCossu2010}
{\sc Hwang, Y. \& Cossu, C.} 2010 Self-sustained process at large scales in
  turbulent channel flow. {\em Phys. Rev. Lett.\/} {\bf 105}, 044505.

\bibitem[Hwang {\em et~al.\/}(2016)Hwang, Willis \&
  Cossu]{HwangWillisCossu2016}
{\sc Hwang, Y., Willis, A.~P. \& Cossu, C.} 2016 Invariant solutions of minimal
  large-scale structures in turbulent channel flow for {$Re_{\tau}$} up to
  1000. {\em J. Fluid Mech.\/} {\bf 802}.

\bibitem[Itano \& Generalis(2009)]{ItanoGeneralis2009}
{\sc Itano, T. \& Generalis, S.~C.} 2009 Hairpin vortex solution in planar
  {C}ouette flow: a tapestry of knotted vortices. {\em Phys. Rev. Lett.\/} {\bf
  102}~(11), 114501.

\bibitem[Itano \& Toh(2001)]{ItanoToh2001}
{\sc Itano, T. \& Toh, S.} 2001 The dynamics of bursting process in wall
  turbulence. {\em J. Phys. Soc. Japan\/} {\bf 70}, 703--716.

\bibitem[Jim\'enez(1987)]{jimenez87}
{\sc Jim\'enez, J.} 1987 Coherent structures and dynamical systems. In {\em
  Proc. CTR Summer School\/}, pp. 323--324. Stanford Univ.

\bibitem[Jim{\'e}nez {\em et~al.\/}(2005)Jim{\'e}nez, Kawahara, Simens, Nagata
  \& Shiba]{JimenezKawaharaSimensNagataShiba2005}
{\sc Jim{\'e}nez, J., Kawahara, G., Simens, M.~P., Nagata, M. \& Shiba, M.}
  2005 Characterization of near-wall turbulence in terms of equilibrium and
  `bursting' solutions. {\em Phys. Fluids\/} {\bf 17}, 015105.

\bibitem[Jim\'enez \& Moin(1991)]{JimenezMoin1991}
{\sc Jim\'enez, J. \& Moin, P.} 1991 The minimal flow unit in near-wall
  turbulence. {\em J. Fluid Mech.\/} {\bf 225}, 213--240.

\bibitem[Kawahara(2005)]{kawahara05}
{\sc Kawahara, G.} 2005 Laminarization of minimal plane {C}ouette flow: {G}oing
  beyond the basin of attraction of turbulence. {\em Phys. Fluids\/} {\bf 17},
  041702.

\bibitem[Kawahara \& Kida(2001)]{KawaharaKida2001}
{\sc Kawahara, G. \& Kida, S.} 2001 Periodic motion embedded in plane {Couette}
  turbulence: regeneration cycle and burst. {\em J. Fluid Mech.\/} {\bf 449},
  291--300.

\bibitem[Kawahara {\em et~al.\/}(2012)Kawahara, Uhlmann \&
  Van~Veen]{KawaharaUhlmannVeen2012}
{\sc Kawahara, G., Uhlmann, M. \& Van~Veen, L.} 2012 The significance of simple
  invariant solutions in turbulent flows. {\em Ann. Rev. of Fluid Mech.\/} {\bf
  44}, 203--225.

\bibitem[Kerswell \& Tutty(2007)]{KerswellTutty2007}
{\sc Kerswell, R.~R. \& Tutty, O.~R.} 2007 Recurrence of travelling waves in
  transitional pipe flow. {\em J. Fluid Mech.\/} {\bf 584}, 69--102.

\bibitem[Kim {\em et~al.\/}(1987)Kim, Moin \& Moser.]{KimMoinMoser1987}
{\sc Kim, J., Moin, P. \& Moser., R.~D.} 1987 Turbulent statistics in fully
  developed channel flow at low {R}eynolds number. {\em J. Fluid Mech.\/} {\bf
  177}, 133--166.

\bibitem[Kreilos \& Eckhardt(2012)]{KreilosEckhardt2012}
{\sc Kreilos, T. \& Eckhardt, B.} 2012 Periodic orbits near onset of chaos in
  plane couette flow. {\em Chaos\/} {\bf 22}~(4), 047505.

\bibitem[Nagata(1990)]{Nagata1990}
{\sc Nagata, M.} 1990 Three-dimensional finite-amplitude solutions in plane
  {Couette} flow: bifurcation from infinity. {\em J. Fluid Mech.\/} {\bf 217},
  519--527.

\bibitem[Park \& Graham(2015)]{ParkGraham2015}
{\sc Park, J.~S. \& Graham, M.~D.} 2015 Exact coherent states and connections
  to turbulent dynamics in minimal channel flow. {\em J. Fluid Mech.\/} {\bf
  782}, 430–454.

\bibitem[Piomelli {\em et~al.\/}(2015)Piomelli, Rouhi \&
  Geurts]{PiomelliRouhiGeurts2015}
{\sc Piomelli, U., Rouhi, A. \& Geurts, B.~J.} 2015 A grid-independent length
  scale for large-eddy simulations. {\em J. Fluid Mech.\/} {\bf 766}, 499--527.

\bibitem[Pope(2000)]{pope00book}
{\sc Pope, S.~B.} 2000 {\em Turbulent flows\/}. Cambridge U. Press.

\bibitem[Pumir(1996)]{Pumir1996}
{\sc Pumir, A.} 1996 Turbulence in homogeneous shear flows. {\em Phys.
  Fluids\/} {\bf 8}, 3112--3127.

\bibitem[Rawat {\em et~al.\/}(2015)Rawat, Cossu, Hwang \&
  Rincon]{RawatCossuHwangRincon2015}
{\sc Rawat, S., Cossu, C., Hwang, Y. \& Rincon, F.} 2015 On the self-sustained
  nature of large-scale motions in turbulent {C}ouette flow. {\em J. Fluid
  Mech.\/} {\bf 782}, 515--540.

\bibitem[Rogers \& Moin(1987)]{RogersMoin1987}
{\sc Rogers, M.~M. \& Moin, P.} 1987 The structure of the vorticity field in
  homogeneous turbulent flows. {\em J. Fluid Mech.\/} {\bf 176}, 33--66.

\bibitem[S{\'a}nchez \& Net(2010)]{SanchezNet2010}
{\sc S{\'a}nchez, J. \& Net, M.} 2010 On the multiple shooting continuation of
  periodic orbits by {N}ewton-{K}rylov methods. {\em J. Bifur. Chaos Appl. Sci.
  Engrg.\/} {\bf 20}, 43--61.

\bibitem[Sasaki {\em et~al.\/}(2016)Sasaki, Kawahara, Sekimoto \&
  Jim{\'e}nez]{SasakiKawaharaSekimotoJimenez2016}
{\sc Sasaki, E., Kawahara, G., Sekimoto, A. \& Jim{\'e}nez, J.} 2016 Unstable
  periodic orbits in plane couette flow with the smagorinsky model. In {\em J.
  Phys.: Conf. Series\/}, , vol. 708, p. 012003. IOP Publishing.

\bibitem[Schmiegel \& Eckhardt(1997)]{SchmiegelEckhardt1997}
{\sc Schmiegel, A. \& Eckhardt, B.} 1997 Fractal stability border in plane
  {C}ouette flow. {\em Phys. Rev. Lett.\/} {\bf 277}, 197--225.

\bibitem[Schneider {\em et~al.\/}(2010)Schneider, Gibson \&
  Burke]{SchneiderGibsonBurke2010}
{\sc Schneider, T.~M., Gibson, J.~F. \& Burke, J.} 2010 Snakes and ladders:
  localized solutions of plane {Couette} flow. {\em Phys. Rev. Lett.\/} {\bf
  104}, 104501.

\bibitem[Scovazzi {\em et~al.\/}(2001)Scovazzi, Jim\'enez \& Moin]{scov2001}
{\sc Scovazzi, G., Jim\'enez, J. \& Moin, P.} 2001 {LES} of the very large
  scales in a ${Re}_\tau=920$ channel. In {\em Proc. Div. Fluid Dyn.\/}, pp.
  KF--5. Am. Phys. Soc.

\bibitem[Sekimoto {\em et~al.\/}(2016)Sekimoto, Dong \&
  Jim\'enez]{SekimotoDongJimenez2016}
{\sc Sekimoto, A., Dong, S. \& Jim\'enez, J.} 2016 Direct numerical simulation
  of statistically stationary and homogeneous shear turbulence and its relation
  to other shear flows. {\em Phys. Fluids\/} {\bf 28}, 035101.

\bibitem[Skufca {\em et~al.\/}(2006)Skufca, J.~A. \&
  Eckhardt]{SkufcaYorkeEckhardt2006}
{\sc Skufca, J.~D., J.~A., Yorke \& Eckhardt, B.} 2006 Edge of chaos in a
  parallel shear flow. {\em Phys. Rev. Lett.\/} {\bf 96}, 174101.

\bibitem[Smagorinsky(1963)]{Smag:63}
{\sc Smagorinsky, J.} 1963 General circulation experiments with the primitive
  equations. {\em Mon. Wea. Rev.\/} {\bf 91}, 99--164.

\bibitem[Tavoularis \& Karnik(1989)]{TavoularisKarnik1989}
{\sc Tavoularis, S. \& Karnik, U.} 1989 Further experiments on the evolution of
  turbulent stresses and scales in uniformly sheared turbulence. {\em J. Fluid
  Mech.\/} {\bf 204}, 457--478.

\bibitem[Toh \& Itano(2003)]{TohItano2003}
{\sc Toh, S. \& Itano, T.} 2003 A periodic-like solution in channel flow. {\em
  J. Fluid Mech.\/} {\bf 481}, 67--76.

\bibitem[Van~Veen {\em et~al.\/}(2011)Van~Veen, Kawahara \&
  Matsumura]{VanVeenGawaharaMatsumura2011}
{\sc Van~Veen, L., Kawahara, G. \& Matsumura, A.} 2011 On matrix-free
  computation of 2d unstable manifolds. {\em SIAM J. SCI. COMPUT.\/} {\bf 33},
  25--44.

\bibitem[van Veen \& Kawahara(2011)]{VanVeenKawahara2011}
{\sc van Veen, L. \& Kawahara, G.} 2011 Homoclinic tangle on the edge of shear
  turbulence. {\em Phys. Rev. Lett.\/} {\bf 107}, 114501.

\bibitem[van Veen {\em et~al.\/}(2006)van Veen, Kida \&
  Kawahara]{VanVeenKidaKawahara2006}
{\sc van Veen, L., Kida, S. \& Kawahara, G.} 2006 Periodic motion representing
  isotropic turbulence. {\em Fluid Dyn. Res.\/} {\bf 38}, 19--46.

\bibitem[Viswanath(2007)]{Viswanath2007}
{\sc Viswanath, D.} 2007 Recurrent motions within plane {Couette} turbulence.
  {\em J. Fluid Mech.\/} {\bf 580}, 339--358.

\bibitem[Viswanath(2009)]{Viswanath2009}
{\sc Viswanath, D.} 2009 The critical layer in pipe flow at high {Reynolds}
  number. {\em Phil. Trans. R. Soc. A\/} {\bf 367}, 561--576.

\bibitem[Waleffe(1997)]{Waleffe1997}
{\sc Waleffe, F.} 1997 On a self-sustaining process in shear flows. {\em Phys.
  Fluids\/} {\bf 9}, 883--900.

\bibitem[Waleffe(2001)]{Waleffe2001}
{\sc Waleffe, F.} 2001 Exact coherent structures in channel flow. {\em J. Fluid
  Mech.\/} {\bf 435}, 93--102.

\bibitem[Wang {\em et~al.\/}(2007)Wang, Gibson \&
  Waleffe]{WangGibsonWaleffe2007}
{\sc Wang, Jue, Gibson, John \& Waleffe, Fabian} 2007 Lower branch coherent
  states in shear flows: Transition and control. {\em Phys. Rev. Lett.\/} {\bf
  98}, 204501.

\bibitem[Wedin \& Kerswell(2004)]{WedinKerswell2004}
{\sc Wedin, H. \& Kerswell, R.} 2004 Exact coherent structures in pipe flow:
  traveling wave solutions. {\em J. Fluid Mech.\/} {\bf 435}, 333--371.

\bibitem[Yasuda {\em et~al.\/}(2014)Yasuda, Goto \&
  Kawahara]{YasudaGotoKawahara2014}
{\sc Yasuda, T., Goto, S. \& Kawahara, G.} 2014 Quasi-cyclic evolution of
  turbulence driven by a steady force in a periodic cube. {\em Fluid Dyn.
  Res.\/} {\bf 46}, 061413.

\bibitem[Zammert \& Eckhardt(2015)]{ZammertEckhardt2015}
{\sc Zammert, S. \& Eckhardt, B.} 2015 Crisis bifurcations in plane
  {P}oiseuille flow. {\em Phys. Rev. E\/} {\bf 91}, 041003.

\end{thebibliography}
